\documentclass[twocolumn,superscriptaddress,prapplied]{revtex4-1}
\usepackage{graphicx}
\usepackage{amssymb}
\usepackage{amsmath}

\usepackage{color}
\usepackage{mathtools}
\usepackage[colorlinks,linkcolor=blue,anchorcolor=blue,citecolor=blue,urlcolor=blue]{hyperref}

\begin{document}
\title{A robust fiber-based quantum thermometer coupled with nitrogen-vacancy centers}
\author{Shao-Chun Zhang}
\author{Yang Dong}
\author{Bo Du}
\author{Hao-Bin Lin}
\author{Shen Li}
\affiliation{{CAS Key Lab of Quantum Information, University of Science and Technology of China, Hefei,
230026, P.R. China}}
\affiliation{{CAS Center For Excellence in Quantum Information and Quantum Physics, University of Science
and Technology of China, Hefei, 230026, P.R. China}}
\author{Wei Zhu}
\author{Guan-Zhong Wang}
\affiliation{Hefei National Laboratory for Physical Science at Microscale, and Department of Physics,
University of Science and Technology of China, Hefei, Anhui 230026, P. R. China}
\author{Xiang-Dong Chen}
\author{Guang-Can Guo}

\author{Fang-Wen Sun}
\email{fwsun@ustc.edu.cn}
\affiliation{{CAS Key Lab of Quantum Information, University of Science and Technology of China, Hefei,
230026, P.R. China}}
\affiliation{{CAS Center For Excellence in Quantum Information and Quantum Physics, University of Science
and Technology of China, Hefei, 230026, P.R. China}}
\date{\today}
\begin{abstract}
The nitrogen-vacancy center in diamond has been broadly applied in quantum sensing since it is sensitive to different physical quantities. Meanwhile, it is difficult to isolate disturbances from unwanted physical quantities in practical applications. Here, we present a robust fiber-based quantum thermometer which can significantly isolate the magnetic field noise and microwave power shift. With a frequency modulation scheme, we realize the temperature measurement by detecting the variation of the sharp-dip in the zero-field optically detected magnetic resonance spectrum in a high-density nitrogen-vacancy ensemble. Thanks to its simplicity and compatibility in implementation and robustness in the isolation of magnetic and microwave noise, this quantum thermometer is then applied to the surface temperature imaging of an electronic chip with a sensitivity of $18$ $\rm{mK}/\sqrt{\rm{Hz}}$. It paves the way to high sensitive temperature measurement in ambiguous environments.
\end{abstract}

\maketitle

\section{Introduction}
The development of a broad range of applications in quantum technologies \cite{obrien2009photonic,awschalom2018quantum}, from quantum networks \cite{duan2010colloquium,reiserer2015cavity} to information processing and quantum sensing \cite{degen2017quantum}, has attracted increasing interests over the past decades. Among them, quantum sensing has been regarded as a new possibility for harnessing the quantum technologies in the real world \cite{degen2017quantum}. In particular, quantum sensor based on nitrogen vacancy (NV) center in diamond is one of the most promising and studied systems for their remarkable optical and spin properties \cite{barry2019sensitivity}. And due to the coupling between NV centers with external perturbations, the sensing functionalities of NV defects are then extended to magnetic field \cite{jensenmagnetometry2017,taylor2008high,li2018enhancing,chen2019superresolution}, electric field \cite{doldeelectricfield2011,michl2019robust}, pressure \cite{lesik2019magnetic,yip2019measuring,hsieh2019imaging} and temperature \cite{chen2011temperature,kucsko2013nanometre,wang2018magnetic} detection, which are usually based on optically detected magnetic resonance (ODMR) with microwave (MW) driving.

Among them of particular interests are the innovative approaches to temperature sensing. A stable and compact thermometer can provide a powerful tool in many areas of physical, chemical, and biological researches \cite{kucsko2013nanometre}. For instance, the recently developed fiber-optic probes coupled with NV centers \cite{liu2013fiber,fedotov2014fiber,blakley2018quantum,dong2018fiber,zhang2019thermal} enabled a compact temperature measurement with a $20$ mK accuracy \cite{fedotov2014fiber}.
However, many of existing methods have the requirement of a bias magnetic field aligning along the NV axis, which make the device complicated and hard for practical measurement.
Moreover, magnetic noise cannot be easily decoupled. Some promising approaches have been studied to overcome this problem, involving, e.g., the techniques with complicated pulse controls \cite{kucsko2013nanometre,fang2013high,wang2015highsensitivity}
and simultaneous tracking of two resonance shifts \cite{wojciechowski2018precision}.
More severely, when a sample contains metal material, the change of the MW power induced by the interaction with the metal material during the temperature scanning has never been considered in previous work \cite{fedotov2014fiber,wojciechowski2018precision}, which highly limits its practical application in ambiguous environments.

Usually, high-density NV ensemble are applied in order to get high sensitivity. In such a sample, a sharp-dip structure around $2.87$ GHz of the zero-field ODMR spectrum \cite{matsuzaki2016optically,mittiga2018imaging,hayashi2018optimization} was recently observed, which can offer a high sensitive temperature sensing. Inspired by Ref. \cite{hayashi2018optimization}, in this Article, we measured this sharp-dip and presented a fiber-based quantum thermometer for practical applications.
Our approach only relied on continuous optical excitation and frequency-modulated MW with the center frequency fixed at the sharp-dip, and subsequently recorded the variation of the sharp-dip through a single lock-in measurement over time. Such a method can decouple the temperature from magnetic field noise and MW power shift during the probing and scanning. Finally, this fiber-based quantum thermometer was successfully applied to the surface temperature imaging of an electronic chip with a temperature sensitivity of $18$ $\rm{mK}/\sqrt{\rm{Hz}}$. This demonstration of robustness against magnetic noise and MW power shift, as well as single lock-in measurement without applying additional magnetic field, makes a significant advance in transiting lab-based systems to practical applications.
\begin{figure*}[t]
\includegraphics[width=0.85\textwidth]{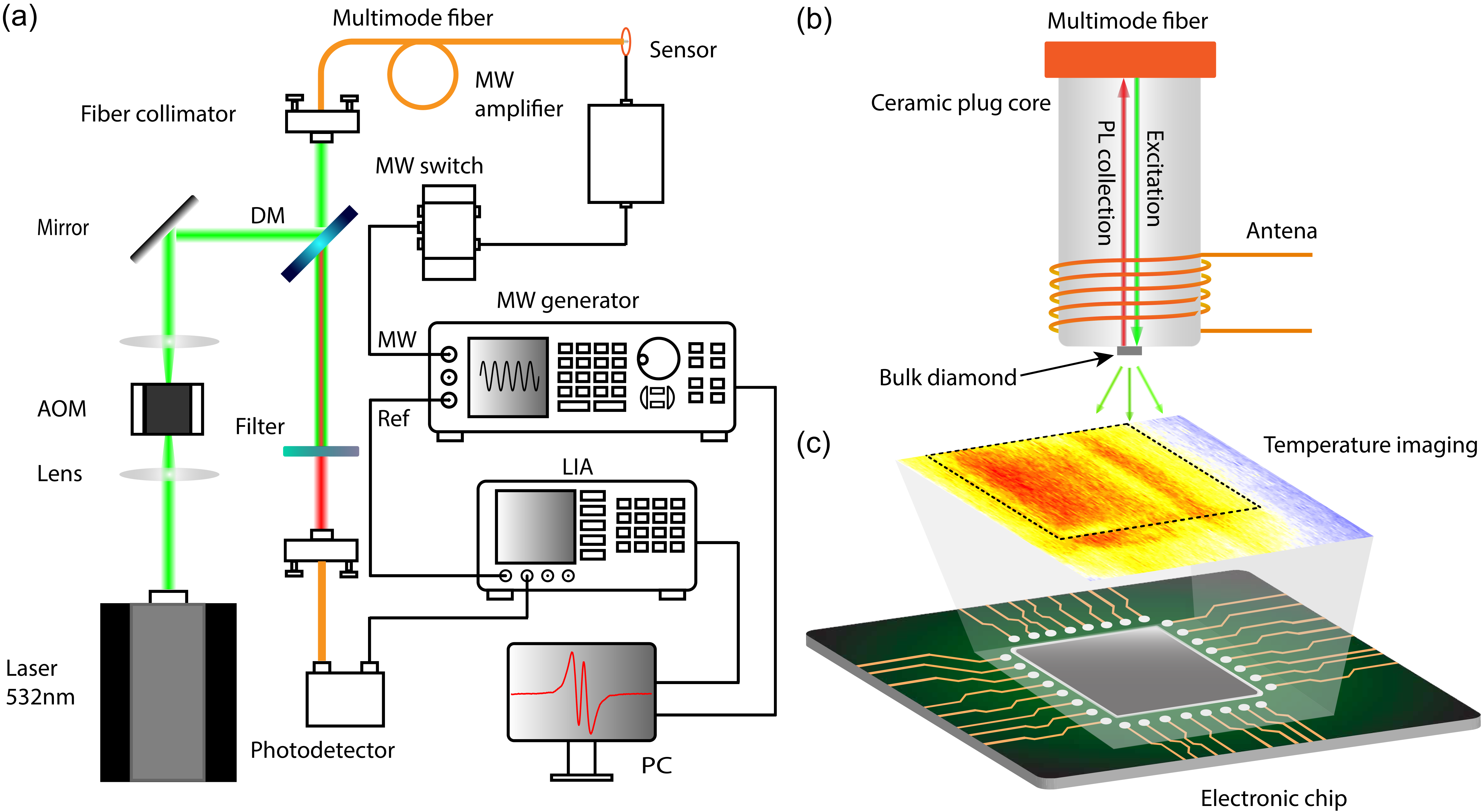}
\caption{\label{experimentsetup}(a) The schematic of hybrid fiber-optical thermometer setup.
DM, long-pass dichroic mirror with the edge wavelength of $658.8$ nm; Ref, the output reference signal from MW generator.
(b) Simplified schematic of the sensor. 
(c) An electronic chip used for temperature imaging via the fiber-based quantum thermometer.}
\end{figure*}

\section{Experimental RESULTS}
\subsection{Experimental setup and conditions for the best sensitivity at room-temperature}
\begin{figure}[t]
\includegraphics[width=0.45\textwidth]{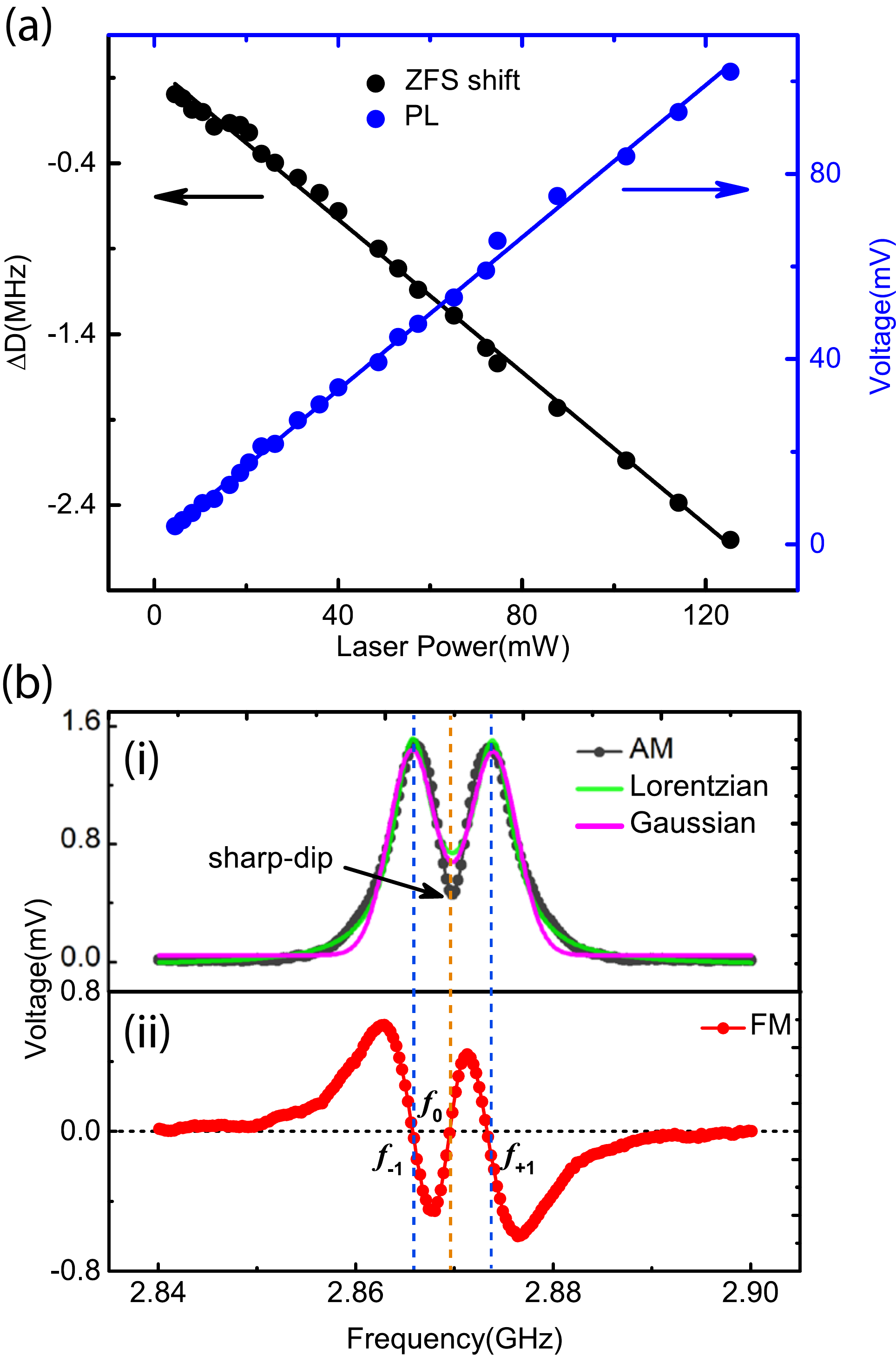}
\caption{\label{fig2}(a) The amount of red fluorescence and ZFS parameter shift ($\Delta$D) as a function of green laser power. Far from saturation, fluorescence increases linearly with laser power plotted as blue line. Due to the laser-heating effect, ZFS parameter decreases linearly plotted as black line. (b) The CW-ODMR spectra recorded with a single MW source by FM and AM without applying a bias magnetic field. (i) The AM spectrum exhibits a sharp-dip structure which cannot be reproduced by either a double Lorentzian (green) or a Gaussian (pink) profile. (ii) The FM spectrum exhibits three typical frequencies of $f_0, f_{\pm{1}}$ corresponding to the FM lock-in signal crossing zero and thus providing the largest temperature response.}
\end{figure}

Usually, conventional confocal scanning system employed in a sensing scheme poses complications in terms of collective control, signal readout and even the stability, leading to the lab-based demonstrations. For practical applications, here, these issues can be addressed by using a homebuilt fiber system to excite and detect the NV centers, as shown in Fig. \ref{experimentsetup}(a).
The NV center ensembles consisted of [$N$] $\approx$ $40$ ppm and [$NV^-$] $\approx$ $0.15$ ppm in diamond with [100] surface orientation grown by plasma assisted chemical vapor deposition. The diamond attached on the tip of a multi-mode optical fiber with a core diameter of $100$ $\mu$m has been mechanically polished and cut into a membrane with dimensions $200 \times$200$\times$100 $\mu$m$^3$, as shown in Fig. \ref{experimentsetup}(b) \cite{zhang2019thermal}. In the experiment, green laser at $532$ nm was sent through an acousto-optical modulator (AOM, AA optoelectronic MT250-A0.5-VIS) and then coupled into the multi-mode optical fiber.
Collected by the same fiber, the photoluminescence (PL) passed through a $647$ nm long pass filter and was finally sent to a photodetector (Thorlabs APD130A2/M). The detected signal was noise filtered and amplified using a lock-in amplifier (LIA, Sine Scientific Instruments OE1022) through either amplitude modulation (AM) or frequency modulation (FM) of MW. The output radio frequency signal of MW generator (Rohde $\&$ Schwarz SMB 100A) was sent to LIA as a reference. Moreover, the output MW was sent through a switch (M-C ZASWA-2-50DR+) to a high-power amplifier (M-C ZHL-16W-43), and finally delivered by a five-turn copper loop with an outer diameter of $0.5$ mm wound around the optical fiber ceramic plug core.

We first studied the amount of red fluorescence from the NV centers as a function of green laser power by processing the AOM with square wave modulation at $333$ Hz (lock-in time constant $\tau$ = $30$ ms), as shown with solid blue dots in Fig. \ref{fig2}(a) together with a linear fit. Far from saturation, fluorescence increases linearly with the laser power. Generally, due to the negligible absorption cross section of single NV centers and inherent power broadening \cite{ahmadi2017enhanced}, reaching a saturated regime becomes more difficult with increasing the ensemble volume and density. In the contrary, the ZFS extracted from the ODMR spectrum decreased as laser power increased, consistent with experimental observations in earlier studies \cite{jensen2013light,fedotov2014fiber,zhang2019thermal}, as shown with solid black dots in Fig. \ref{fig2}(a). Normally, diamond thermometers with higher power pump are expected to offer better sensitivity. However, for the fiber-based thermometer, the laser-heating effect resulted from high power pump can significantly affect the detection accuracy of temperature. Thus it is necessary to set the laser power below $10$ mW. In this case, the temperature of the diamond can be kept at room temperature, and the local temperature variation can be transferred to the diamond \cite{zhang2019thermal} and detected with NV centers.

\begin{figure*}[t]
\includegraphics[width=0.9\textwidth]{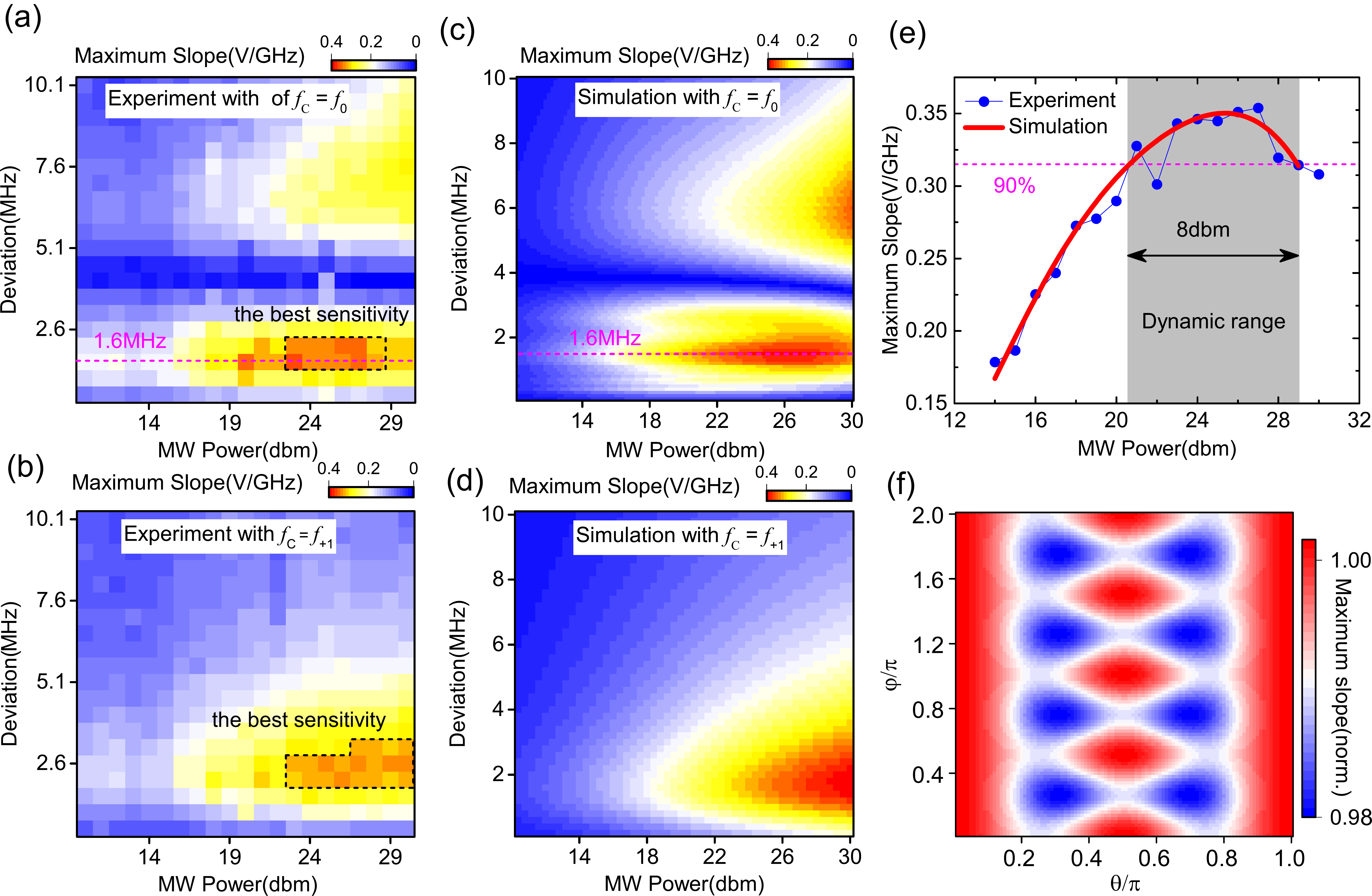}
\caption{\label{fig3} (a-b) The maximum slope of FM ODMR spectrum as a function of MW power and modulation deviation with the center frequency fixed at $f_0$ and $f_{+1}$, respectively. (c-d) Simulation result with Eq. (\ref{cwodmr}) and Eq. (\ref{fmodmr}) compared with experimental results (a-b). Simulation parameters are $T_1=7.1$ ms, $T_2^*=0.32$ $\mu$s, $T_2=15.4$ $\mu$s. The optical readout pump rate is $\Gamma_p=3$ MHz. (e) Cross sections of the pink dashed lines in (a) and (c). The sensitivity variation less than $10\%$ in the shaded area suggests the dynamic range of MW power shift. (f) Simulation of the maximum slope of ODMR spectra with $f_0$ as a function of the vector geomagnetic field. Here, the extracted maximum slopes are normalized, and the simulation parameters are identical with (c-d).}
\end{figure*}

The fiber-based diamond thermometer is based on the technique of ODMR. The NV center consists of a substitutional nitrogen atom combining with a vacancy in a neighboring lattice site of the diamond crystal. In the absence of external magnetic field, its spin triplet ground state is degenerated by spin-spin interaction into a singlet state $\rm{m_s=0}$ and a doublet $\rm{m_s=\pm{1}}$, separated by the temperature-dependent ZFS parameter $D_{gs}$ with $\frac{\partial{D_{gs}}}{\partial{T}}\approx74$ $\rm{Hz}/\rm{mK}$ \cite{acosta2010temperature,chen2011temperature,li2017temperature}. The doublet $\rm{m_s=\pm{1}}$ will split due to the static magnetic field along the NV axis. However, for high-density NV ensemble, a qualitatively distinct spectrum can be observed, consisting of a pair of resonances centered at $D_{gs}$, mainly due to the local strain from internal defect of crystal \cite{noauthor2013nitrogen,matsuzaki2016optically,mittiga2018imaging,hayashi2018optimization}, via the Hamiltonian of ground state:
\begin{align}\label{Hamiltonian}
H=&(D_{gs}(T)+\Pi_{z})S_{z}^{2}+(\delta{B_{z}}+A_{zz}I_{z})S_{z} \nonumber\\
&-\Pi_{x}(S_{x}^{2}-S_{y}^{2})+\Pi_{y}(S_{x}S_{y}+S_{y}S_{x})\nonumber\\
&+\Omega\cos(2\pi{ft})S_{x} \text{.}
\end{align}
Here, $\hat{z}$ is the NV axis. $\hat{x}$ is defined such that one of the carbon-vacancy bonds lies in the $x-z$ plane. $\vec{S}$ and $\vec{I}$ are the electronic spin-$1$ operators of NV and nuclear spin-$1$ operators of the host $^{14}$N. $\delta{B_z}$ represents the axial magnetic field and $A_{zz}=2.16$ MHz. The terms $\Pi_{\{x,y\}}$ characterize the coupling between NV and strain-electric field \cite{doherty2012theory}. $\Omega$ and $f$ are the MW amplitude and frequency, respectively. Moreover, the continuous-wave (CW) ODMR spectra of NV ensembles can be simulated using the steady-state solution of a five-level Bloch equation \cite{ahmadi2017enhanced,ella2017optimised}, where the expression can be defined as a sum of Lorentzian distributions:
\begin{align}\label{cwodmr}
S_{CW}(f)=&1-\sum^{N}_{i=1}\sum^{1}_{m_I=-1}\Big \{\frac{\mathcal{C}_i\gamma_i^2}{(f-f_{-,i}-m_{I}A_{zz})^2+\gamma_i^2}\nonumber\\
&+\frac{\mathcal{C}_i\gamma_i^2}{(f-f_{+,i}-m_{I}A_{zz})^2+\gamma_i^2}\Big \} \text{,}
\end{align}
with $f_{\pm,i}=D_{gs}(T)+\Pi_{z,i}\pm\sqrt{\Pi_{x,i}^2+\Pi_{y,i}^2+\delta B_{z,i}^2}$.
Here the contrast $\mathcal{C}$ as well as the half width at half maximum $\gamma$ are both $\Omega$-dependent.

In more recent studies, lock-in techniques were feasibly used to continuously monitor the ODMR spectrum. In this case, small resonance frequency shifts were detected by applying pre-calibrated scale factor \cite{clevenson2018robust} determined by the optical pump power, MW power and even the detection efficiency \cite{jensen2013light}. By sweeping the frequency of MW with a time-independent amplitude or frequency, both AM and FM are currently used to ODMR detection \cite{schoenfeld2011real,shao2016diamond,ella2017optimised,ahmadi2017enhanced,wojciechowski2018precision,schloss2018simultaneous,clevenson2018robust}.

Here, by applying a sinusoidal-wave AM of MW with a $333$ Hz modulation rate and a $100\%$ modulation depth in the absence of bias magnetic field, we obtained an ODMR spectrum as plotted in Fig. \ref{fig2}(b)(i).
The line shape of each resonance cannot be captured by either a Gaussian or Lorentzian profile, leading a sharp-dip structure around $2.87$ GHz. However,
the model of the randomized magnetic field $\delta B_z$ and electric-strain field $\Pi_{x,y,z}$ in Eq. (\ref{Hamiltonian}) has succeeded in fitting the ODMR spectra \cite{matsuzaki2016optically,hayashi2018optimization,mittiga2018imaging}. 
Despite one can determine the temperature by fitting the ODMR curves to obtain the transition frequencies, the actual temperature measurement is usually carried out by fixing the MW frequency and recording the variation of LIA signal in time.

In order to monitor the LIA signal response to the environmental fluctuation, we employed the FM of MW and then amplified the resulting ac signal using LIA. The applied MW frequency varied as $F(t)\approx f_{\rm{c}}+f_{\rm{d}}\rm{cos}(2{\pi}\emph{f}_{\rm{mod}}t)$ where $f_{\rm{c}}$, $f_{\rm{d}}$, and $f_{\rm{mod}}$ are center frequency, deviation and modulation frequency, respectively. By sweeping the MW center frequency, the FM ODMR spectrum extracted from Eq. (\ref{cwodmr}) can be defined as:
\begin{align}\label{fmodmr}
U(f_c)=\int_{0}^{1/f_{\rm{mod}}} &\mathcal{A}S_{CW}(F(t))\cos(2\pi{f_{\rm{mod}}}t)\, dt \text{,}
\end{align}
where $\mathcal{A}$ is proportional to the lock-in amplifier gain factor. The FM ODMR spectrum from the experiment is shown in Fig. \ref{fig2}(b)(ii), which is almost proportional to the derivative of the AM signal in Fig. \ref{fig2}(b)(i). Thus the signals with frequencies of $f_{-1}$, $f_{0}$, and $f_{+1}$ crossing zero in Fig. \ref{fig2}(b)(ii) correspond to the extreme points of AM ODMR spectrum in Fig. \ref{fig2}(b)(i). In particular, $f_{0}\approx D_{gs}(T)+\Pi_{z,eff}$, where $\Pi_{z,eff}$ is the effective axial strain which causes the shift of the overall spectrum. Such three frequencies also provide the largest temperature response.

\begin{figure*}[t]
\includegraphics[width=0.75\textwidth]{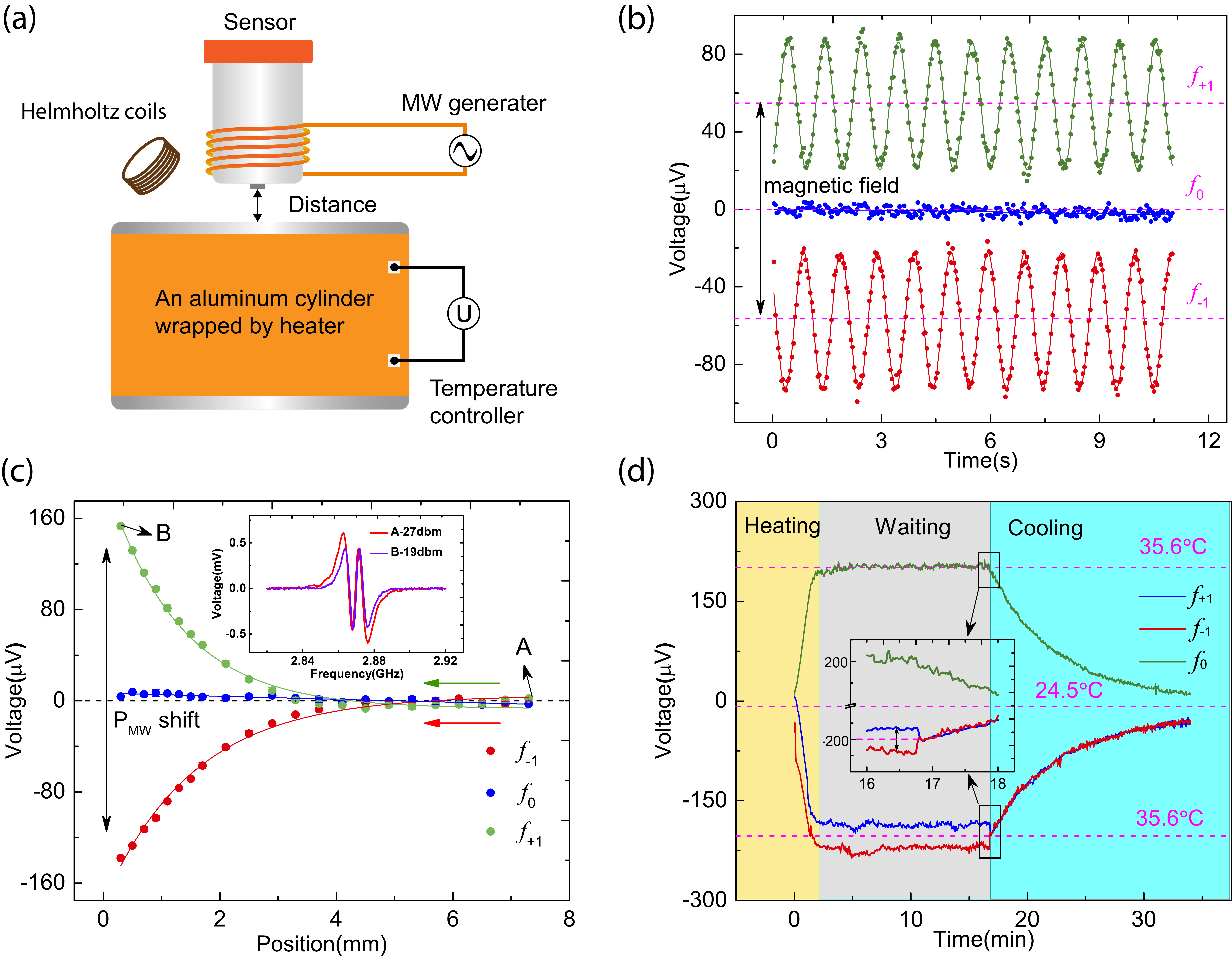}
\caption{\label{fig4} Isolation of magnetic field noise and MW power shift by FM scheme with the center frequency fixed at $f_0$.
(a) Simplified schematic of the experimental setup. An aluminum cylinder wrapped by heater is close to the sensor without touching, and the distance between them can be adjusted. A Helmholtz coils provides additional test magnetic field.
(b) LIA signal with a $5.2$ $\mu$T bias magnetic field oscillating at $1$ Hz via a Helmholtz coils.
(c) LIA signal response as a function of the distance from the sensor to the aluminum cylinder.
(d) LIA signal with temperature at three processes including heating, waiting and cooling. A jump occurred at $17$ min (inset) mainly due to the disappearance of static magnetic field when the heater was powered off.}
\end{figure*}
The temperature sensitivity using ODMR spectrum is limited by the inverse of the maximum MW-dependent rate of fluorescence change, which is given by \cite{hayashi2018optimization}:
\begin{equation}\label{Sensitivity}
\partial{T_{\rm{min}}}\sim\frac{1}{\Big(\frac{\partial{D_{gs}}}{\partial{T}}\Big)^{-1}\Big{\vert}\frac{\partial{U}}{\partial{f}}\Big{\vert}_{\rm{max}}} \text{,}
\end{equation}
where $U$ is the ODMR voltage from the NV centers. Obviously, the temperature sensitivity is directly proportional to $[|\partial{U}/\partial{f}|_{\rm{max}}]^{-1}$ which can be determined by optimizing the MW power P$_{\rm{MW}}$ and modulation deviation $f_{\rm{d}}$. With the center frequency fixed at $f_0$ and $f_{+1}$, we extracted the maximum slope $|\partial{U}/\partial{f}|_{\rm{max}}$ of the ODMR spectra as a function of MW power $P_{MW}$ and modulation deviation as shown in Fig. \ref{fig3}(a-b). It shows excellent agreement with the simulations via Eq. (\ref{cwodmr}) and Eq. (\ref{fmodmr}) in Fig. \ref{fig3}(c-d).
The black dotted boxes in Fig. \ref{fig2}(a) and Fig. \ref{fig2}(b) indicate the conditions, consisting of the MW power range from $22-27$ dbm and the modulation deviation range from $1.6-2$ MHz, offering the best temperature sensitivity. Fig. \ref{fig3}(e) depicts the cross sections of the pink dashed lines in Fig. \ref{fig3}(a) and Fig. \ref{fig3}(c), and the gray area demonstrates a $8$ dbm dynamic range for MW power shift, where the temperature sensitivity varies less than $10\%$. Generally, the existence of the Earth's magnetic field (about $0.05$ mT) always effects the ODMR spectrum even in the absence of additional magnetic field. By numerical simulation with optimized $f_d=1.6$ MHz, $P_{MW}=26$ dbm, and identical parameters in Fig. \ref{fig3}(c-d), we extracted the maximum slope of ODMR spectrum as a function of vector geomagnetic field shown in Fig. \ref{fig3}(f), where $\theta$ and $\phi$ are the polar and azimuth angles with respect to [100] and [010] crystal orientation of the diamond, respectively. Critically, the slope variation is not more than $2\%$, indicating that the effect of the Earth's magnetic field is negligibly small. Thus the sensitivity remains constant.

\subsection{Isolation of magnetic field noise and MW power shift}
In the sensing processes, due to coupling with magnetic field, MW and temperature, the LIA signal variation $S$ has an approximately linear spectral dependence, which can be expressed as
\begin{equation}\label{Signal Response1}
S\sim\sum^{N}_{i=1}\Big \lgroup\frac{\partial{U_i}}{\partial{B_i}}\Big{\vert}_f\Delta B_i+\frac{\partial{U_i}}{\partial{P_i}}\Big{\vert}_f\Delta P_i+\frac{\partial{U_i}}{\partial{T}}\Big{\vert}_f\Delta T\Big \rgroup \text{,}
\end{equation}
where $U_i$ is the ODMR voltage from each single NV center, $\Delta$ denotes the change in the parameters, $B_i$ and $P_i$ represent the static magnetic field and MW field projected onto each NV center.
Therefore, the NV center spin resonance frequencies are sensitive to the magnetic field, in which case magnetic field variations can no longer be neglected. To experimentally demonstrate the isolation of magnetic field noise with single lock-in measurement, we finely set the MW center frequency until the LIA signal was zero with an optimized $f_d$ and $P_{\rm{MW}}$, and then applied a $5.2$ $\mathrm{\mu}$T test field oscillating at $1$ Hz via a Helmholtz coils, as shown in Fig. \ref{fig4}(a). The LIA responses over time with the center frequencies fixed at $f_0$ and $f_{\pm1}$ were recorded ($f_{mod}=3.3$ KHz, $\tau=30$ ms), as shown in Fig. \ref{fig4}(b). The LIA signal with the center frequency of $f_{\pm1}$ oscillated out-phase at $1$ Hz but the signal with $f_0$ remained unchanged, demonstrating that the thermometer with the center frequency of $f_0$ was unaffected by magnetic noise.
Thus, for weak magnetic noise, the term $\frac{\partial{U_i}}{\partial{B}}\Big{\vert}_{f_0}\Delta B_i=0$ in Eq. \eqref{Signal Response1} can be neglected.

\begin{figure*}
\includegraphics[width=0.85\textwidth]{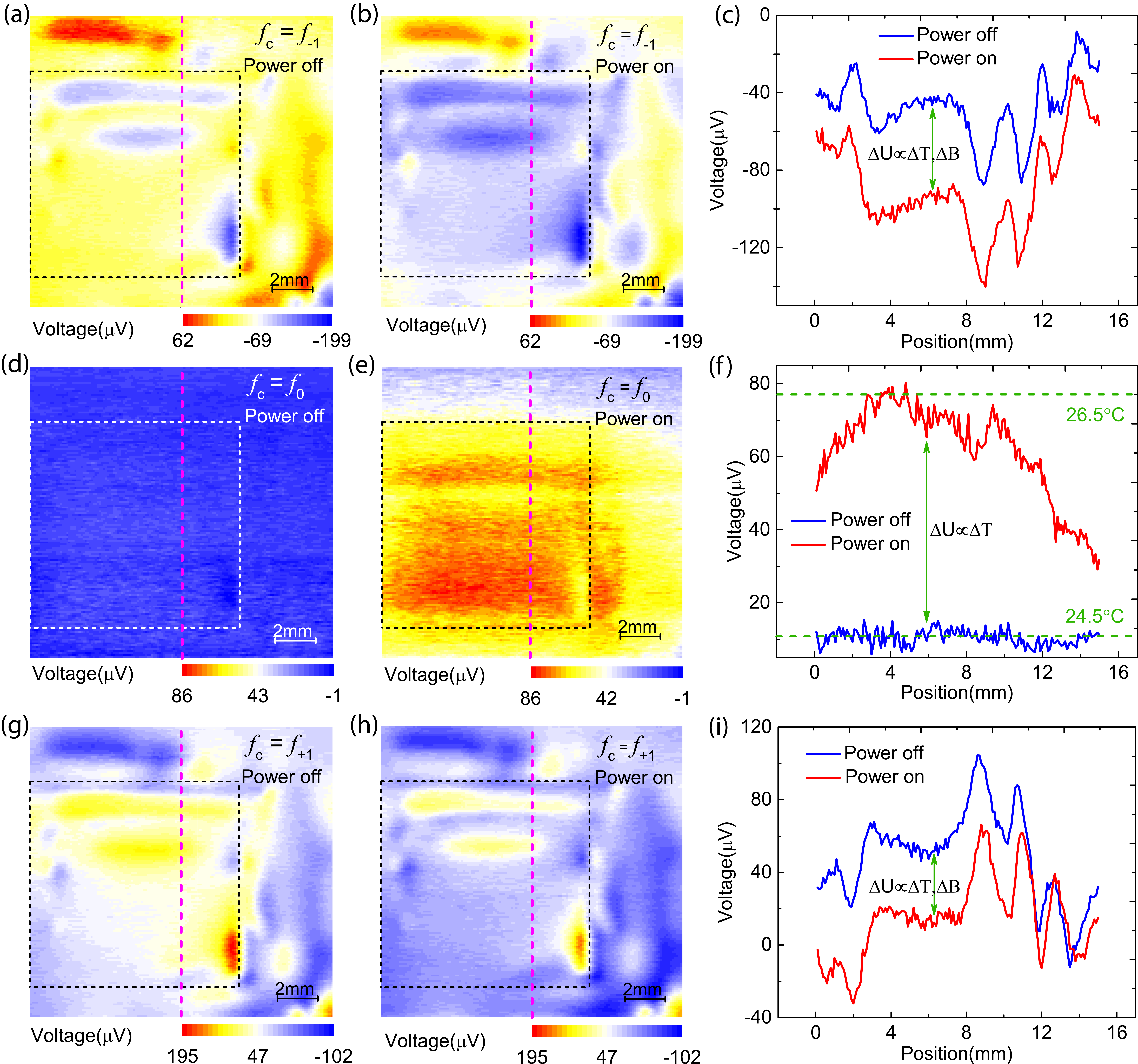}
\caption{\label{fig5} Temperature scanning of an electronic chip (dashed box).
(a-b) Recording the LIA signal with center frequencies fixed at $f_{-1}$ when the chip was powered off and on respectively. (c) Cross sections of the pink dashed lines in (a-b).
(d-e) Recording the LIA signal with center frequencies fixed at $f_0$ when the chip was powered off and on respectively. (f) Cross sections of the pink dashed lines in (d-e).
(g-h) Recording the LIA signal with center frequencies fixed at $f_{+1}$ when the chip was powered off and on respectively. (i) Cross sections of the pink dashed lines in (g-h).}
\end{figure*}

On the other hand, the test object, such as metal and other dielectric, near the sensor will affect the MW radiation and further the contrast of ODMR spectrum, which make these previous multi-resonance schemes \cite{kitazawa2017vector,wojciechowski2018precision,schloss2018simultaneous,yahata2019demonstration} ineffective. In this case, to simulate MW power shift during the temperature probing, we used an aluminum cylinder as a test object as shown in Fig. \ref{fig4}(a). Firstly, we moved the sensor away from the aluminum (distance between each other was $8$ mm) and then finely set the MW center frequency until the LIA signal was zero. By contrast, Fig. \ref{fig4}(c) shows the experimental LIA signal response as a function of the sensor position with the center frequencies fixed at $f_0$ and $f_{\pm}$ , together with exponential fit plotted by solid lines. Based on the ODMR spectra as shown in the inset of Fig. \ref{fig4}(c), the distance from A to B accompanied the reduction of MW power felt by the sensor approximately from $27$ to $19$ dbm. Critically, the LIA signal with center frequency of $f_{\pm1}$ changed out-phase while the signal of $f_0$ did not change.
Indeed, MW power shift, such as the case at point B, will change the sensitivity shown in Fig. \ref{fig3}(a-b), and further affect the accuracy of temperature measurement.
However, our scheme has a tolerance for MW power shift, as shown in Fig. \ref{fig4}(c).
Thus for small MW power shift, the term $\frac{\partial{U_i}}{\partial{P}}\Big{\vert}_{f_0}\Delta P_i$ can also be neglected. Similarly, the Eq. \eqref{Signal Response1} can be simplified as
\begin{equation}\label{Signal Response2}
S\sim\sum^{4}_{i=1}\Big \lgroup\frac{\partial{U_i}}{\partial{T}}\Big{\vert}_{f_0}\Delta T\Big \rgroup \text{.}
\end{equation}

To this end, the magnetic field noise and MW power shift can be significantly isolated via fixing the center frequency at $f_0$. We next performed the temperature measurement and confirmed its robustness. The same aluminum cylinder was heated by a resistive heater (Thorlabs, HT10K) and its temperature was detected by the fiber-based quantum thermometer, as shown in Fig. \ref{fig4}(a). Firstly, we moved the thermometer close to the aluminum without touching, and then finely set the MW center frequency until the LIA signal was zero. By contrast, the LIA signals were recorded at three processes including heating ($0-2$ min), waiting ($2-17$ min) and cooling ($17-29$ min), as shown in Fig. \ref{fig4}(d). With heating, the temperature of the aluminum increased from room temperature of $24.5^{\circ}\rm{C}$ to $35.6^{\circ}\rm{C}$, and cooled down to room temperature after a thermostatic process. In particular, LIA signal of $f_0$ (green line) changes out-phase compared with that of $f_{\pm1}$ (red and blue lines) as expected. Moreover, an opposite shift occurred between the LIA signal of $f_{-1}$ (red plot) and $f_{+1}$ (blue plot) when the heater was powered on, but disappeared after the heater was powered off, indicating that a weak static magnetic field was generated during the heating and waiting process, as shown in the inset of Fig. \ref{fig4}(d). However, averaging such two LIA signals of $f_{+1}$ and $f_{-1}$ can not only give almost the same amplitude as LIA signal of $f_0$, but also cancel the effect of magnetic field noise and MW power shift. This agrees with the result of previous multi-frequency modulation \cite{wojciechowski2018precision}. By comparison, our scheme can measure the temperature by just a single-frequency modulation with the same sensitivity, as well as a wide dynamic range.

\subsection{Surface temperature imaging of an electronic chip}
Recently, temperature imaging has been an important research method. In particular, temperature imaging of electronic chips attracts much attention in the field of electronic engineering. Chip temperature imaging can not only reflect the working state, but also the energy-consumption of the chip. However, the temperature measurement module in the chip, which offers a low accuracy of around $200$ mK, can only provide its temperature for average, rather than the distribution. Here, our fiber-based quantum thermometer was successfully applied to image the surface temperature of a chip, thereby proving the practicability and robustness of our method.

As shown in Fig. \ref{experimentsetup}(c), we fixed an electronic chip (Raspherry Pi Zero W) on the translation stage. Similarly, we first moved the sensor away from the chip, then finely set the MW center frequency until the output LIA signal was zero, and finally moved the sensor close to the chip without touching. In this case, LIA can completely record the surface temperature, magnetic field as well as the metal-induced MW power shift. By scanning, the LIA signal images with the center frequency of $f_{-1}$ with the chip power off and on are shown in Fig. \ref{fig5}(a-b). Both images exhibited almost the same features even the chip was powered off, indicating that the effect of MW power shift was dominant during the scanning. Moreover, the cross sections depicted in Fig. \ref{fig5}(c) demonstrated an overall shift $\Delta\rm{U}$ between such two images, which was resulted from the magnetic field change $\Delta B$ and temperature variation $\Delta$T. Subsequently, we fixed the center frequency at $f_0$ and performed the scanning again. Such a process has suppressed the MW power shift as well as magnetic field noise, leading to an uniform feature in Fig. \ref{fig5}(d) when the chip was powered off. Hence, we can reconstruct the surface temperature distribution directly from Fig. \ref{fig5}(e) when the chip was powered on. The cross sections depicted in Fig. \ref{fig5}(f) suggested that the chip exhibited a maximum temperature of $26.5^{\circ}$C when power was on and an uniform temperature of $24.5^{\circ}$C when power was off. In addition, compared with Fig. \ref{fig5}(a-c), the LIA signal images with the center frequency fixed at $f_{+1}$ not only showed the out-phase variation result from MW power shift, but also the in-phase variation result from temperature change, as shown in Fig. \ref{fig5}(g-i).

\begin{figure}
\includegraphics[width=0.47\textwidth]{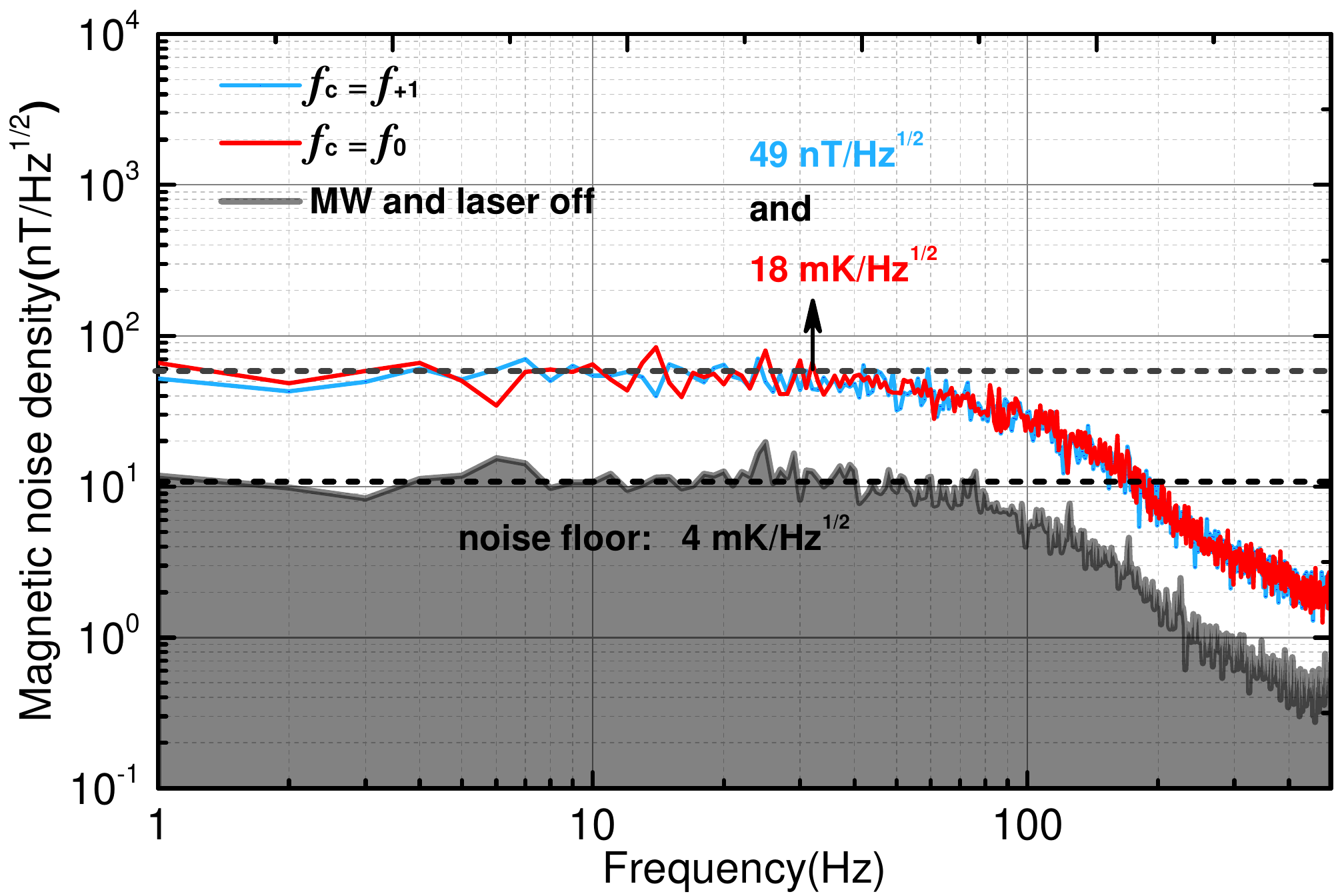}
\caption{\label{fig6}Plots of the magnetic noise spectral densities with the center frequencies fixed at $f_{+1}$ (light blue) and $f_{0}$ (red), as well as a reference spectrum obtained with both of MW and laser turned off (gray). Each plot was obtained by recording the time traces of the lock-in output with the length of $10$ s, then dividing such $10$ s data set into ten $1$ s segments, and finally averaging the absolute value of the Fourier transform of each segment. All the recordings were at conditions of the best sensitivity as mentioned in Fig. \ref{fig2}.}
\end{figure}
\subsection{Temperature sensitivity}
In order to characterize the sensitivity of the fiber-based quantum thermometer, we measured the the spectral noise density. By finely setting the MW center frequency until the LIA signal crossed zero thus providing the largest temperature response, we then continuously recorded the LIA signal of $f_0$  and $f_{+1}$  for $10$ s with a $1$ kHz sampling, and calculated noise spectra of $f_0$  and $f_{+1}$, as shown in Fig. \ref{fig6}. Both of such two spectra showed the same response to the environmental variation. However, the LIA signal of $f_{+1}$ and $f_0$ have been proved sensitive to magnetic field and temperature respectively. Hence, in the configuration of magnetic field sensing, the spectrum of $f_{+1}$ gives a magnetic sensitivity of $49$ $\rm{nT}/\sqrt{\rm{Hz}}$. For temperature sensing, LIA signal of $f_0$ had no response to magnetic noise but achieved a temperature sensitivity of 18 $\rm{mK}/\sqrt{\rm{Hz}}$. For reference, a spectrum obtained with the MW and laser turned off is depicted in Fig. \ref{fig6} with the gray plot, which demonstrates a noise floor of $4$ $\rm{mK}/\sqrt{\rm{Hz}}$ limited by the lock-in input noise and the detected shot noise.

\section{DISCUSSION and CONCLUSION}
Generally, the size of the diamond attached on the fiber tip not only determines the spatial resolution of temperature imaging, but also the time for the sensor to reach thermal equilibrium. In the current implementation, the sensor consists of a $200 \times200$ $\mu$m$^2$ bulk diamond, optical fiber ceramic plug core with $2.4$ mm diameter and copper wires wrapped around it. These large size makes the system take about $100$ ms to reach thermal equilibrium, which seriously reduces the temperature imaging rate. Using micron-sized particles and miniaturizing the sensor will indeed solve these problems, but these advantages are usually brought at the expense of sensitivity. A challenge is thus to increase the density of NV centers while maintaining good spin coherence properties.

Our fiber-based quantum thermometer used a bulk diamond processed by a standard annealing treatments, which provided a sensitivity of $18$ mK/$\sqrt{\rm{Hz}}$ without optimized fluorescence excitation and collection. Since the measurement was performed at room-temperature, the requirement of low laser power limited the sensitivity enhancement. However, by electron irradiation treatment, the fast electrons can knock carbon atoms out of the lattice sites producing vacancies and interstitial carbons \cite{farfurnik2017enhanced, mclellan2016patterned,eichhorn2019optimizing}, leading to a high probability for NV center combination and at least $20$ times enhancement to the density of NV centers \cite{zhang2019thermal}. On the other hand, recent study with a matched micro-concave mirror to a sphered optic-fiber end has achieved over $25$ times more fluorescence collection from NV enriched micrometer-sized diamond \cite{duan2018enhancing}.  All of these methods can boost temperature sensitivity toward sub-$1$ mK $/\sqrt{\rm{Hz}}$.


In conclusion, we have presented a robust fiber-based thermometer coupled with NV centers at room-temperature, allowing a sensitivity of $18$ $\rm{mK}/\sqrt{\rm{Hz}}$.
Such a method can protect the temperature measurement from the environmental magnetic noise and MW power shift with a single lock-in measurement. With the fiber-based thermometer, we have successfully imaged the surface temperature distribution of an electronic chip. Thanks to its simplicity and robustness, such a quantum thermometer paves a significant step towards practical applications in microscale thermal detection for integrated chips and biology endoscopes with high precision in ambiguous environments.
\section*{Acknowledgment}
This work is supported by the National Key Research and Development Program of China (No. 2017YFA0304504), the National Natural Science
Foundation of China (Nos. 91536219, and 91850102), Anhui Initiative in Quantum Information
Technologies (No. AHY130000), the Science Challenge Project (Grant
No. TZ2018003).
\nocite{*}

\begin{thebibliography}{46}%
\makeatletter
\providecommand \@ifxundefined [1]{%
 \@ifx{#1\undefined}
}%
\providecommand \@ifnum [1]{%
 \ifnum #1\expandafter \@firstoftwo
 \else \expandafter \@secondoftwo
 \fi
}%
\providecommand \@ifx [1]{%
 \ifx #1\expandafter \@firstoftwo
 \else \expandafter \@secondoftwo
 \fi
}%
\providecommand \natexlab [1]{#1}%
\providecommand \enquote  [1]{``#1''}%
\providecommand \bibnamefont  [1]{#1}%
\providecommand \bibfnamefont [1]{#1}%
\providecommand \citenamefont [1]{#1}%
\providecommand \href@noop [0]{\@secondoftwo}%
\providecommand \href [0]{\begingroup \@sanitize@url \@href}%
\providecommand \@href[1]{\@@startlink{#1}\@@href}%
\providecommand \@@href[1]{\endgroup#1\@@endlink}%
\providecommand \@sanitize@url [0]{\catcode `\\12\catcode `\$12\catcode
  `\&12\catcode `\#12\catcode `\^12\catcode `\_12\catcode `\%12\relax}%
\providecommand \@@startlink[1]{}%
\providecommand \@@endlink[0]{}%
\providecommand \url  [0]{\begingroup\@sanitize@url \@url }%
\providecommand \@url [1]{\endgroup\@href {#1}{\urlprefix }}%
\providecommand \urlprefix  [0]{URL }%
\providecommand \Eprint [0]{\href }%
\providecommand \doibase [0]{http://dx.doi.org/}%
\providecommand \selectlanguage [0]{\@gobble}%
\providecommand \bibinfo  [0]{\@secondoftwo}%
\providecommand \bibfield  [0]{\@secondoftwo}%
\providecommand \translation [1]{[#1]}%
\providecommand \BibitemOpen [0]{}%
\providecommand \bibitemStop [0]{}%
\providecommand \bibitemNoStop [0]{.\EOS\space}%
\providecommand \EOS [0]{\spacefactor3000\relax}%
\providecommand \BibitemShut  [1]{\csname bibitem#1\endcsname}%
\let\auto@bib@innerbib\@empty
\bibitem [{\citenamefont {O'Brien}\ \emph {et~al.}(2009)\citenamefont
  {O'Brien}, \citenamefont {Furusawa},\ and\ \citenamefont
  {Vuckovic}}]{obrien2009photonic}%
  \BibitemOpen
  \bibfield  {author} {\bibinfo {author} {\bibfnamefont {J.~L.}\ \bibnamefont
  {O'Brien}}, \bibinfo {author} {\bibfnamefont {A.}~\bibnamefont {Furusawa}}, \
  and\ \bibinfo {author} {\bibfnamefont {J.}~\bibnamefont {Vuckovic}},\
  }\bibfield  {title} {\enquote {\bibinfo {title} {Photonic quantum
  technologies},}\ }\href {\doibase 10.1038/nphoton.2009.229} {\bibfield
  {journal} {\bibinfo  {journal} {Nat. Photon.}\ }\textbf {\bibinfo {volume}
  {3}},\ \bibinfo {pages} {687} (\bibinfo {year} {2009})}\BibitemShut {NoStop}%
\bibitem [{\citenamefont {Awschalom}\ \emph {et~al.}(2018)\citenamefont
  {Awschalom}, \citenamefont {Hanson}, \citenamefont {Wrachtrup},\ and\
  \citenamefont {Zhou}}]{awschalom2018quantum}%
  \BibitemOpen
  \bibfield  {author} {\bibinfo {author} {\bibfnamefont {D.~D.}\ \bibnamefont
  {Awschalom}}, \bibinfo {author} {\bibfnamefont {R.}~\bibnamefont {Hanson}},
  \bibinfo {author} {\bibfnamefont {J.}~\bibnamefont {Wrachtrup}}, \ and\
  \bibinfo {author} {\bibfnamefont {B.~B.}\ \bibnamefont {Zhou}},\ }\bibfield
  {title} {\enquote {\bibinfo {title} {Quantum technologies with optically
  interfaced solid-state spins},}\ }\href {\doibase 10.1038/s41566-018-0232-2}
  {\bibfield  {journal} {\bibinfo  {journal} {Nat. Photon.}\ }\textbf {\bibinfo
  {volume} {12}},\ \bibinfo {pages} {516} (\bibinfo {year} {2018})}\BibitemShut
  {NoStop}%
\bibitem [{\citenamefont {Duan}\ and\ \citenamefont
  {Monroe}(2010)}]{duan2010colloquium}%
  \BibitemOpen
  \bibfield  {author} {\bibinfo {author} {\bibfnamefont {L.-M.}\ \bibnamefont
  {Duan}}\ and\ \bibinfo {author} {\bibfnamefont {C.}~\bibnamefont {Monroe}},\
  }\bibfield  {title} {\enquote {\bibinfo {title} {Colloquium: Quantum networks
  with trapped ions},}\ }\href {\doibase 10.1103/RevModPhys.82.1209} {\bibfield
   {journal} {\bibinfo  {journal} {Rev. Mod. Phys.}\ }\textbf {\bibinfo
  {volume} {82}},\ \bibinfo {pages} {1209} (\bibinfo {year}
  {2010})}\BibitemShut {NoStop}%
\bibitem [{\citenamefont {Reiserer}\ and\ \citenamefont
  {Rempe}(2015)}]{reiserer2015cavity}%
  \BibitemOpen
  \bibfield  {author} {\bibinfo {author} {\bibfnamefont {A.}~\bibnamefont
  {Reiserer}}\ and\ \bibinfo {author} {\bibfnamefont {G.}~\bibnamefont
  {Rempe}},\ }\bibfield  {title} {\enquote {\bibinfo {title} {Cavity-based
  quantum networks with single atoms and optical photons},}\ }\href {\doibase
  10.1103/RevModPhys.87.1379} {\bibfield  {journal} {\bibinfo  {journal} {Rev.
  Mod. Phys.}\ }\textbf {\bibinfo {volume} {87}},\ \bibinfo {pages} {1379}
  (\bibinfo {year} {2015})}\BibitemShut {NoStop}%
\bibitem [{\citenamefont {Degen}\ \emph {et~al.}(2017)\citenamefont {Degen},
  \citenamefont {Reinhard},\ and\ \citenamefont
  {Cappellaro}}]{degen2017quantum}%
  \BibitemOpen
  \bibfield  {author} {\bibinfo {author} {\bibfnamefont {C.~L.}\ \bibnamefont
  {Degen}}, \bibinfo {author} {\bibfnamefont {F.}~\bibnamefont {Reinhard}}, \
  and\ \bibinfo {author} {\bibfnamefont {P.}~\bibnamefont {Cappellaro}},\
  }\bibfield  {title} {\enquote {\bibinfo {title} {Quantum sensing},}\ }\href
  {\doibase 10.1103/RevModPhys.89.035002} {\bibfield  {journal} {\bibinfo
  {journal} {Rev. Mod. Phys.}\ }\textbf {\bibinfo {volume} {89}},\ \bibinfo
  {pages} {035002} (\bibinfo {year} {2017})}\BibitemShut {NoStop}%
\bibitem [{\citenamefont {Barry}\ \emph {et~al.}(2019)\citenamefont {Barry},
  \citenamefont {Schloss}, \citenamefont {Bauch}, \citenamefont {Turner},
  \citenamefont {Hart}, \citenamefont {Pham},\ and\ \citenamefont
  {Walsworth}}]{barry2019sensitivity}%
  \BibitemOpen
  \bibfield  {author} {\bibinfo {author} {\bibfnamefont {J.~F.}\ \bibnamefont
  {Barry}}, \bibinfo {author} {\bibfnamefont {J.~M.}\ \bibnamefont {Schloss}},
  \bibinfo {author} {\bibfnamefont {E.}~\bibnamefont {Bauch}}, \bibinfo
  {author} {\bibfnamefont {M.~J.}\ \bibnamefont {Turner}}, \bibinfo {author}
  {\bibfnamefont {C.~A.}\ \bibnamefont {Hart}}, \bibinfo {author}
  {\bibfnamefont {L.~M.}\ \bibnamefont {Pham}}, \ and\ \bibinfo {author}
  {\bibfnamefont {R.~L.}\ \bibnamefont {Walsworth}},\ }\bibfield  {title}
  {\enquote {\bibinfo {title} {Sensitivity optimization for {NV}-diamond
  magnetometry},}\ }\href {http://arxiv.org/abs/1903.08176} {\bibfield
  {journal} {\bibinfo  {journal} {{arXiv}:1903.08176}\ } (\bibinfo {year}
  {2019})}\BibitemShut {NoStop}%
\bibitem [{\citenamefont {Jensen}\ \emph {et~al.}(2017)\citenamefont {Jensen},
  \citenamefont {Kehayias},\ and\ \citenamefont
  {Budker}}]{jensenmagnetometry2017}%
  \BibitemOpen
  \bibfield  {author} {\bibinfo {author} {\bibfnamefont {K.}~\bibnamefont
  {Jensen}}, \bibinfo {author} {\bibfnamefont {P.}~\bibnamefont {Kehayias}}, \
  and\ \bibinfo {author} {\bibfnamefont {D.}~\bibnamefont {Budker}},\
  }\bibfield  {title} {\enquote {\bibinfo {title} {Magnetometry with
  nitrogen-vacancy centers in diamond},}\ }in\ \href
  {https://link.springer.com/chapter/10.1007/978-3-319-34070-8_18} {\emph
  {\bibinfo {booktitle} {High sensitivity magnetometers}}}\ (\bibinfo
  {publisher} {Springer},\ \bibinfo {year} {2017})\ pp.\ \bibinfo {pages}
  {553--576}\BibitemShut {NoStop}%
\bibitem [{\citenamefont {Taylor}\ \emph {et~al.}(2008)\citenamefont {Taylor},
  \citenamefont {Cappellaro}, \citenamefont {Childress}, \citenamefont {Jiang},
  \citenamefont {Budker}, \citenamefont {Hemmer}, \citenamefont {Yacoby},
  \citenamefont {Walsworth},\ and\ \citenamefont {Lukin}}]{taylor2008high}%
  \BibitemOpen
  \bibfield  {author} {\bibinfo {author} {\bibfnamefont {J.}~\bibnamefont
  {Taylor}}, \bibinfo {author} {\bibfnamefont {P.}~\bibnamefont {Cappellaro}},
  \bibinfo {author} {\bibfnamefont {L.}~\bibnamefont {Childress}}, \bibinfo
  {author} {\bibfnamefont {L.}~\bibnamefont {Jiang}}, \bibinfo {author}
  {\bibfnamefont {D.}~\bibnamefont {Budker}}, \bibinfo {author} {\bibfnamefont
  {P.}~\bibnamefont {Hemmer}}, \bibinfo {author} {\bibfnamefont
  {A.}~\bibnamefont {Yacoby}}, \bibinfo {author} {\bibfnamefont
  {R.}~\bibnamefont {Walsworth}}, \ and\ \bibinfo {author} {\bibfnamefont
  {M.}~\bibnamefont {Lukin}},\ }\bibfield  {title} {\enquote {\bibinfo {title}
  {High-sensitivity diamond magnetometer with nanoscale resolution},}\ }\href
  {https://aip.scitation.org/doi/10.1063/1.4949357} {\bibfield  {journal}
  {\bibinfo  {journal} {Nat. Phys.}\ }\textbf {\bibinfo {volume} {4}},\
  \bibinfo {pages} {810} (\bibinfo {year} {2008})}\BibitemShut {NoStop}%
\bibitem [{\citenamefont {Li}\ \emph {et~al.}(2018)\citenamefont {Li},
  \citenamefont {Dong}, \citenamefont {Xu}, \citenamefont {Li}, \citenamefont
  {Chen}, \citenamefont {Du}, \citenamefont {Ge}, \citenamefont {Guo},\ and\
  \citenamefont {Sun}}]{li2018enhancing}%
  \BibitemOpen
  \bibfield  {author} {\bibinfo {author} {\bibfnamefont {C.-H.}\ \bibnamefont
  {Li}}, \bibinfo {author} {\bibfnamefont {Y.}~\bibnamefont {Dong}}, \bibinfo
  {author} {\bibfnamefont {J.-Y.}\ \bibnamefont {Xu}}, \bibinfo {author}
  {\bibfnamefont {D.-F.}\ \bibnamefont {Li}}, \bibinfo {author} {\bibfnamefont
  {X.-D.}\ \bibnamefont {Chen}}, \bibinfo {author} {\bibfnamefont {A.~M.}\
  \bibnamefont {Du}}, \bibinfo {author} {\bibfnamefont {Y.-S.}\ \bibnamefont
  {Ge}}, \bibinfo {author} {\bibfnamefont {G.-C.}\ \bibnamefont {Guo}}, \ and\
  \bibinfo {author} {\bibfnamefont {F.-W.}\ \bibnamefont {Sun}},\ }\bibfield
  {title} {\enquote {\bibinfo {title} {Enhancing the sensitivity of a single
  electron spin sensor by multi-frequency control},}\ }\href {\doibase
  10.1063/1.5042796} {\bibfield  {journal} {\bibinfo  {journal} {Appl. Phys.
  Lett.}\ }\textbf {\bibinfo {volume} {113}},\ \bibinfo {pages} {072401}
  (\bibinfo {year} {2018})}\BibitemShut {NoStop}%
\bibitem [{\citenamefont {Chen}\ \emph {et~al.}(2019)\citenamefont {Chen},
  \citenamefont {Li}, \citenamefont {Zheng}, \citenamefont {Li}, \citenamefont
  {Du}, \citenamefont {Dong}, \citenamefont {Dong}, \citenamefont {Guo},\ and\
  \citenamefont {Sun}}]{chen2019superresolution}%
  \BibitemOpen
  \bibfield  {author} {\bibinfo {author} {\bibfnamefont {X.-D.}\ \bibnamefont
  {Chen}}, \bibinfo {author} {\bibfnamefont {D.-F.}\ \bibnamefont {Li}},
  \bibinfo {author} {\bibfnamefont {Y.}~\bibnamefont {Zheng}}, \bibinfo
  {author} {\bibfnamefont {S.}~\bibnamefont {Li}}, \bibinfo {author}
  {\bibfnamefont {B.}~\bibnamefont {Du}}, \bibinfo {author} {\bibfnamefont
  {Y.}~\bibnamefont {Dong}}, \bibinfo {author} {\bibfnamefont {C.-H.}\
  \bibnamefont {Dong}}, \bibinfo {author} {\bibfnamefont {G.-C.}\ \bibnamefont
  {Guo}}, \ and\ \bibinfo {author} {\bibfnamefont {F.-W.}\ \bibnamefont
  {Sun}},\ }\bibfield  {title} {\enquote {\bibinfo {title} {Superresolution
  multifunctional sensing with the nitrogen-vacancy center in diamond},}\
  }\href {\doibase 10.1103/PhysRevApplied.12.044039} {\bibfield  {journal}
  {\bibinfo  {journal} {Phys. Rev. Appl.}\ }\textbf {\bibinfo {volume} {12}},\
  \bibinfo {pages} {044039} (\bibinfo {year} {2019})}\BibitemShut {NoStop}%
\bibitem [{\citenamefont {Dolde}\ \emph {et~al.}(2011)\citenamefont {Dolde},
  \citenamefont {Fedder}, \citenamefont {Doherty}, \citenamefont {N{\"o}bauer},
  \citenamefont {Rempp}, \citenamefont {Balasubramanian}, \citenamefont {Wolf},
  \citenamefont {Reinhard}, \citenamefont {Hollenberg}, \citenamefont {Jelezko}
  \emph {et~al.}}]{doldeelectricfield2011}%
  \BibitemOpen
  \bibfield  {author} {\bibinfo {author} {\bibfnamefont {F.}~\bibnamefont
  {Dolde}}, \bibinfo {author} {\bibfnamefont {H.}~\bibnamefont {Fedder}},
  \bibinfo {author} {\bibfnamefont {M.~W.}\ \bibnamefont {Doherty}}, \bibinfo
  {author} {\bibfnamefont {T.}~\bibnamefont {N{\"o}bauer}}, \bibinfo {author}
  {\bibfnamefont {F.}~\bibnamefont {Rempp}}, \bibinfo {author} {\bibfnamefont
  {G.}~\bibnamefont {Balasubramanian}}, \bibinfo {author} {\bibfnamefont
  {T.}~\bibnamefont {Wolf}}, \bibinfo {author} {\bibfnamefont {F.}~\bibnamefont
  {Reinhard}}, \bibinfo {author} {\bibfnamefont {L.~C.}\ \bibnamefont
  {Hollenberg}}, \bibinfo {author} {\bibfnamefont {F.}~\bibnamefont {Jelezko}},
   \emph {et~al.},\ }\bibfield  {title} {\enquote {\bibinfo {title}
  {Electric-field sensing using single diamond spins},}\ }\href
  {https://www.nature.com/articles/nphys1969} {\bibfield  {journal} {\bibinfo
  {journal} {Nat. Phys.}\ }\textbf {\bibinfo {volume} {7}},\ \bibinfo {pages}
  {459} (\bibinfo {year} {2011})}\BibitemShut {NoStop}%
\bibitem [{\citenamefont {Michl}\ \emph {et~al.}(2019)\citenamefont {Michl},
  \citenamefont {Steiner}, \citenamefont {Denisenko}, \citenamefont
  {B\"{u}lau}, \citenamefont {Zimmermann}, \citenamefont {Nakamura},
  \citenamefont {Sumiya}, \citenamefont {Onoda}, \citenamefont {Neumann},
  \citenamefont {Isoya},\ and\ \citenamefont {Wrachtrup}}]{michl2019robust}%
  \BibitemOpen
  \bibfield  {author} {\bibinfo {author} {\bibfnamefont {J.}~\bibnamefont
  {Michl}}, \bibinfo {author} {\bibfnamefont {J.}~\bibnamefont {Steiner}},
  \bibinfo {author} {\bibfnamefont {A.}~\bibnamefont {Denisenko}}, \bibinfo
  {author} {\bibfnamefont {A.}~\bibnamefont {B\"{u}lau}}, \bibinfo {author}
  {\bibfnamefont {A.}~\bibnamefont {Zimmermann}}, \bibinfo {author}
  {\bibfnamefont {K.}~\bibnamefont {Nakamura}}, \bibinfo {author}
  {\bibfnamefont {H.}~\bibnamefont {Sumiya}}, \bibinfo {author} {\bibfnamefont
  {S.}~\bibnamefont {Onoda}}, \bibinfo {author} {\bibfnamefont
  {P.}~\bibnamefont {Neumann}}, \bibinfo {author} {\bibfnamefont
  {J.}~\bibnamefont {Isoya}}, \ and\ \bibinfo {author} {\bibfnamefont
  {J.}~\bibnamefont {Wrachtrup}},\ }\bibfield  {title} {\enquote {\bibinfo
  {title} {Robust and accurate electric field sensing with solid state spin
  ensembles},}\ }\href {\doibase 10.1021/acs.nanolett.9b00900} {\bibfield
  {journal} {\bibinfo  {journal} {Nano Lett.}\ }\textbf {\bibinfo {volume}
  {19}},\ \bibinfo {pages} {4904} (\bibinfo {year} {2019})}\BibitemShut
  {NoStop}%
\bibitem [{\citenamefont {Lesik}\ \emph {et~al.}(2019)\citenamefont {Lesik},
  \citenamefont {Plisson}, \citenamefont {Toraille}, \citenamefont {Renaud},
  \citenamefont {Occelli}, \citenamefont {Schmidt}, \citenamefont {Salord},
  \citenamefont {Delobbe}, \citenamefont {Debuisschert}, \citenamefont
  {Rondin}, \citenamefont {Loubeyre},\ and\ \citenamefont
  {Roch}}]{lesik2019magnetic}%
  \BibitemOpen
  \bibfield  {author} {\bibinfo {author} {\bibfnamefont {M.}~\bibnamefont
  {Lesik}}, \bibinfo {author} {\bibfnamefont {T.}~\bibnamefont {Plisson}},
  \bibinfo {author} {\bibfnamefont {L.}~\bibnamefont {Toraille}}, \bibinfo
  {author} {\bibfnamefont {J.}~\bibnamefont {Renaud}}, \bibinfo {author}
  {\bibfnamefont {F.}~\bibnamefont {Occelli}}, \bibinfo {author} {\bibfnamefont
  {M.}~\bibnamefont {Schmidt}}, \bibinfo {author} {\bibfnamefont
  {O.}~\bibnamefont {Salord}}, \bibinfo {author} {\bibfnamefont
  {A.}~\bibnamefont {Delobbe}}, \bibinfo {author} {\bibfnamefont
  {T.}~\bibnamefont {Debuisschert}}, \bibinfo {author} {\bibfnamefont
  {L.}~\bibnamefont {Rondin}}, \bibinfo {author} {\bibfnamefont
  {P.}~\bibnamefont {Loubeyre}}, \ and\ \bibinfo {author} {\bibfnamefont
  {J.-F.}\ \bibnamefont {Roch}},\ }\bibfield  {title} {\enquote {\bibinfo
  {title} {Magnetic measurements on micrometer-sized samples under high
  pressure using designed {NV} centers},}\ }\href {\doibase
  10.1126/science.aaw4329} {\bibfield  {journal} {\bibinfo  {journal}
  {Science}\ }\textbf {\bibinfo {volume} {366}},\ \bibinfo {pages} {1359}
  (\bibinfo {year} {2019})}\BibitemShut {NoStop}%
\bibitem [{\citenamefont {Yip}\ \emph {et~al.}(2019)\citenamefont {Yip},
  \citenamefont {Ho}, \citenamefont {Yu}, \citenamefont {Chen}, \citenamefont
  {Zhang}, \citenamefont {Kasahara}, \citenamefont {Mizukami}, \citenamefont
  {Shibauchi}, \citenamefont {Matsuda}, \citenamefont {Goh},\ and\
  \citenamefont {Yang}}]{yip2019measuring}%
  \BibitemOpen
  \bibfield  {author} {\bibinfo {author} {\bibfnamefont {K.~Y.}\ \bibnamefont
  {Yip}}, \bibinfo {author} {\bibfnamefont {K.~O.}\ \bibnamefont {Ho}},
  \bibinfo {author} {\bibfnamefont {K.~Y.}\ \bibnamefont {Yu}}, \bibinfo
  {author} {\bibfnamefont {Y.}~\bibnamefont {Chen}}, \bibinfo {author}
  {\bibfnamefont {W.}~\bibnamefont {Zhang}}, \bibinfo {author} {\bibfnamefont
  {S.}~\bibnamefont {Kasahara}}, \bibinfo {author} {\bibfnamefont
  {Y.}~\bibnamefont {Mizukami}}, \bibinfo {author} {\bibfnamefont
  {T.}~\bibnamefont {Shibauchi}}, \bibinfo {author} {\bibfnamefont
  {Y.}~\bibnamefont {Matsuda}}, \bibinfo {author} {\bibfnamefont {S.~K.}\
  \bibnamefont {Goh}}, \ and\ \bibinfo {author} {\bibfnamefont
  {S.}~\bibnamefont {Yang}},\ }\bibfield  {title} {\enquote {\bibinfo {title}
  {Measuring magnetic field texture in correlated electron systems under
  extreme conditions},}\ }\href {\doibase 10.1126/science.aaw4278} {\bibfield
  {journal} {\bibinfo  {journal} {Science}\ }\textbf {\bibinfo {volume}
  {366}},\ \bibinfo {pages} {1355} (\bibinfo {year} {2019})}\BibitemShut
  {NoStop}%
\bibitem [{\citenamefont {Hsieh}\ \emph {et~al.}(2019)\citenamefont {Hsieh},
  \citenamefont {Bhattacharyya}, \citenamefont {Zu}, \citenamefont {Mittiga},
  \citenamefont {Smart}, \citenamefont {Machado}, \citenamefont {Kobrin},
  \citenamefont {Höhn}, \citenamefont {Rui}, \citenamefont {Kamrani},
  \citenamefont {Chatterjee}, \citenamefont {Choi}, \citenamefont {Zaletel},
  \citenamefont {Struzhkin}, \citenamefont {Moore}, \citenamefont {Levitas},
  \citenamefont {Jeanloz},\ and\ \citenamefont {Yao}}]{hsieh2019imaging}%
  \BibitemOpen
  \bibfield  {author} {\bibinfo {author} {\bibfnamefont {S.}~\bibnamefont
  {Hsieh}}, \bibinfo {author} {\bibfnamefont {P.}~\bibnamefont
  {Bhattacharyya}}, \bibinfo {author} {\bibfnamefont {C.}~\bibnamefont {Zu}},
  \bibinfo {author} {\bibfnamefont {T.}~\bibnamefont {Mittiga}}, \bibinfo
  {author} {\bibfnamefont {T.~J.}\ \bibnamefont {Smart}}, \bibinfo {author}
  {\bibfnamefont {F.}~\bibnamefont {Machado}}, \bibinfo {author} {\bibfnamefont
  {B.}~\bibnamefont {Kobrin}}, \bibinfo {author} {\bibfnamefont {T.~O.}\
  \bibnamefont {Höhn}}, \bibinfo {author} {\bibfnamefont {N.~Z.}\ \bibnamefont
  {Rui}}, \bibinfo {author} {\bibfnamefont {M.}~\bibnamefont {Kamrani}},
  \bibinfo {author} {\bibfnamefont {S.}~\bibnamefont {Chatterjee}}, \bibinfo
  {author} {\bibfnamefont {S.}~\bibnamefont {Choi}}, \bibinfo {author}
  {\bibfnamefont {M.}~\bibnamefont {Zaletel}}, \bibinfo {author} {\bibfnamefont
  {V.~V.}\ \bibnamefont {Struzhkin}}, \bibinfo {author} {\bibfnamefont {J.~E.}\
  \bibnamefont {Moore}}, \bibinfo {author} {\bibfnamefont {V.~I.}\ \bibnamefont
  {Levitas}}, \bibinfo {author} {\bibfnamefont {R.}~\bibnamefont {Jeanloz}}, \
  and\ \bibinfo {author} {\bibfnamefont {N.~Y.}\ \bibnamefont {Yao}},\
  }\bibfield  {title} {\enquote {\bibinfo {title} {Imaging stress and magnetism
  at high pressures using a nanoscale quantum sensor},}\ }\href {\doibase
  10.1126/science.aaw4352} {\bibfield  {journal} {\bibinfo  {journal}
  {Science}\ }\textbf {\bibinfo {volume} {366}},\ \bibinfo {pages} {1349}
  (\bibinfo {year} {2019})}\BibitemShut {NoStop}%
\bibitem [{\citenamefont {Chen}\ \emph {et~al.}(2011)\citenamefont {Chen},
  \citenamefont {Dong}, \citenamefont {Sun}, \citenamefont {Zou}, \citenamefont
  {Cui}, \citenamefont {Han},\ and\ \citenamefont {Guo}}]{chen2011temperature}%
  \BibitemOpen
  \bibfield  {author} {\bibinfo {author} {\bibfnamefont {X.-D.}\ \bibnamefont
  {Chen}}, \bibinfo {author} {\bibfnamefont {C.-H.}\ \bibnamefont {Dong}},
  \bibinfo {author} {\bibfnamefont {F.-W.}\ \bibnamefont {Sun}}, \bibinfo
  {author} {\bibfnamefont {C.-L.}\ \bibnamefont {Zou}}, \bibinfo {author}
  {\bibfnamefont {J.-M.}\ \bibnamefont {Cui}}, \bibinfo {author} {\bibfnamefont
  {Z.-F.}\ \bibnamefont {Han}}, \ and\ \bibinfo {author} {\bibfnamefont
  {G.-C.}\ \bibnamefont {Guo}},\ }\bibfield  {title} {\enquote {\bibinfo
  {title} {Temperature dependent energy level shifts of nitrogen-vacancy
  centers in diamond},}\ }\href {\doibase 10.1063/1.3652910} {\bibfield
  {journal} {\bibinfo  {journal} {Appl. Phys. Lett.}\ }\textbf {\bibinfo
  {volume} {99}},\ \bibinfo {pages} {161903} (\bibinfo {year}
  {2011})}\BibitemShut {NoStop}%
\bibitem [{\citenamefont {Kucsko}\ \emph {et~al.}(2013)\citenamefont {Kucsko},
  \citenamefont {Maurer}, \citenamefont {Yao}, \citenamefont {Kubo},
  \citenamefont {Noh}, \citenamefont {Lo}, \citenamefont {Park},\ and\
  \citenamefont {Lukin}}]{kucsko2013nanometre}%
  \BibitemOpen
  \bibfield  {author} {\bibinfo {author} {\bibfnamefont {G.}~\bibnamefont
  {Kucsko}}, \bibinfo {author} {\bibfnamefont {P.~C.}\ \bibnamefont {Maurer}},
  \bibinfo {author} {\bibfnamefont {N.~Y.}\ \bibnamefont {Yao}}, \bibinfo
  {author} {\bibfnamefont {M.}~\bibnamefont {Kubo}}, \bibinfo {author}
  {\bibfnamefont {H.~J.}\ \bibnamefont {Noh}}, \bibinfo {author} {\bibfnamefont
  {P.~K.}\ \bibnamefont {Lo}}, \bibinfo {author} {\bibfnamefont
  {H.}~\bibnamefont {Park}}, \ and\ \bibinfo {author} {\bibfnamefont {M.~D.}\
  \bibnamefont {Lukin}},\ }\bibfield  {title} {\enquote {\bibinfo {title}
  {Nanometre-scale thermometry in a living cell},}\ }\href {\doibase
  10.1038/nature12373} {\bibfield  {journal} {\bibinfo  {journal} {Nature}\
  }\textbf {\bibinfo {volume} {500}},\ \bibinfo {pages} {54} (\bibinfo {year}
  {2013})}\BibitemShut {NoStop}%
\bibitem [{\citenamefont {Wang}\ \emph {et~al.}(2018)\citenamefont {Wang},
  \citenamefont {Liu}, \citenamefont {Leong}, \citenamefont {Zeng},
  \citenamefont {Feng}, \citenamefont {Li}, \citenamefont {Dolde},
  \citenamefont {Fedder}, \citenamefont {Wrachtrup}, \citenamefont {Cui},
  \citenamefont {Yang}, \citenamefont {Li},\ and\ \citenamefont
  {Liu}}]{wang2018magnetic}%
  \BibitemOpen
  \bibfield  {author} {\bibinfo {author} {\bibfnamefont {N.}~\bibnamefont
  {Wang}}, \bibinfo {author} {\bibfnamefont {G.-Q.}\ \bibnamefont {Liu}},
  \bibinfo {author} {\bibfnamefont {W.-H.}\ \bibnamefont {Leong}}, \bibinfo
  {author} {\bibfnamefont {H.}~\bibnamefont {Zeng}}, \bibinfo {author}
  {\bibfnamefont {X.}~\bibnamefont {Feng}}, \bibinfo {author} {\bibfnamefont
  {S.-H.}\ \bibnamefont {Li}}, \bibinfo {author} {\bibfnamefont
  {F.}~\bibnamefont {Dolde}}, \bibinfo {author} {\bibfnamefont
  {H.}~\bibnamefont {Fedder}}, \bibinfo {author} {\bibfnamefont
  {J.}~\bibnamefont {Wrachtrup}}, \bibinfo {author} {\bibfnamefont {X.-D.}\
  \bibnamefont {Cui}}, \bibinfo {author} {\bibfnamefont {S.}~\bibnamefont
  {Yang}}, \bibinfo {author} {\bibfnamefont {Q.}~\bibnamefont {Li}}, \ and\
  \bibinfo {author} {\bibfnamefont {R.-B.}\ \bibnamefont {Liu}},\ }\bibfield
  {title} {\enquote {\bibinfo {title} {Magnetic criticality enhanced hybrid
  nanodiamond thermometer under ambient conditions},}\ }\href {\doibase
  10.1103/PhysRevX.8.011042} {\bibfield  {journal} {\bibinfo  {journal} {Phys.
  Rev. X}\ }\textbf {\bibinfo {volume} {8}},\ \bibinfo {pages} {011042}
  (\bibinfo {year} {2018})}\BibitemShut {NoStop}%
\bibitem [{\citenamefont {Liu}\ \emph {et~al.}(2013)\citenamefont {Liu},
  \citenamefont {Cui}, \citenamefont {Sun}, \citenamefont {Song}, \citenamefont
  {Feng}, \citenamefont {Wang}, \citenamefont {Zhu}, \citenamefont {Lou},\ and\
  \citenamefont {Wang}}]{liu2013fiber}%
  \BibitemOpen
  \bibfield  {author} {\bibinfo {author} {\bibfnamefont {X.}~\bibnamefont
  {Liu}}, \bibinfo {author} {\bibfnamefont {J.}~\bibnamefont {Cui}}, \bibinfo
  {author} {\bibfnamefont {F.}~\bibnamefont {Sun}}, \bibinfo {author}
  {\bibfnamefont {X.}~\bibnamefont {Song}}, \bibinfo {author} {\bibfnamefont
  {F.}~\bibnamefont {Feng}}, \bibinfo {author} {\bibfnamefont {J.}~\bibnamefont
  {Wang}}, \bibinfo {author} {\bibfnamefont {W.}~\bibnamefont {Zhu}}, \bibinfo
  {author} {\bibfnamefont {L.}~\bibnamefont {Lou}}, \ and\ \bibinfo {author}
  {\bibfnamefont {G.}~\bibnamefont {Wang}},\ }\bibfield  {title} {\enquote
  {\bibinfo {title} {Fiber-integrated diamond-based magnetometer},}\ }\href
  {\doibase 10.1063/1.4823548} {\bibfield  {journal} {\bibinfo  {journal}
  {Appl. Phys. Lett.}\ }\textbf {\bibinfo {volume} {103}},\ \bibinfo {pages}
  {143105} (\bibinfo {year} {2013})}\BibitemShut {NoStop}%
\bibitem [{\citenamefont {Fedotov}\ \emph {et~al.}(2014)\citenamefont
  {Fedotov}, \citenamefont {Blakley}, \citenamefont {Serebryannikov},
  \citenamefont {Safronov}, \citenamefont {Velichansky}, \citenamefont
  {Scully},\ and\ \citenamefont {Zheltikov}}]{fedotov2014fiber}%
  \BibitemOpen
  \bibfield  {author} {\bibinfo {author} {\bibfnamefont {I.~V.}\ \bibnamefont
  {Fedotov}}, \bibinfo {author} {\bibfnamefont {S.}~\bibnamefont {Blakley}},
  \bibinfo {author} {\bibfnamefont {E.~E.}\ \bibnamefont {Serebryannikov}},
  \bibinfo {author} {\bibfnamefont {N.~A.}\ \bibnamefont {Safronov}}, \bibinfo
  {author} {\bibfnamefont {V.~L.}\ \bibnamefont {Velichansky}}, \bibinfo
  {author} {\bibfnamefont {M.~O.}\ \bibnamefont {Scully}}, \ and\ \bibinfo
  {author} {\bibfnamefont {A.~M.}\ \bibnamefont {Zheltikov}},\ }\bibfield
  {title} {\enquote {\bibinfo {title} {Fiber-based thermometry using optically
  detected magnetic resonance},}\ }\href {\doibase 10.1063/1.4904798}
  {\bibfield  {journal} {\bibinfo  {journal} {Appl. Phys. Lett.}\ }\textbf
  {\bibinfo {volume} {105}},\ \bibinfo {pages} {261109} (\bibinfo {year}
  {2014})}\BibitemShut {NoStop}%
\bibitem [{\citenamefont {Blakley}\ \emph {et~al.}(2018)\citenamefont
  {Blakley}, \citenamefont {Fedotov}, \citenamefont {Becker},\ and\
  \citenamefont {Zheltikov}}]{blakley2018quantum}%
  \BibitemOpen
  \bibfield  {author} {\bibinfo {author} {\bibfnamefont {S.~M.}\ \bibnamefont
  {Blakley}}, \bibinfo {author} {\bibfnamefont {I.~V.}\ \bibnamefont
  {Fedotov}}, \bibinfo {author} {\bibfnamefont {J.}~\bibnamefont {Becker}}, \
  and\ \bibinfo {author} {\bibfnamefont {A.~M.}\ \bibnamefont {Zheltikov}},\
  }\bibfield  {title} {\enquote {\bibinfo {title} {Quantum stereomagnetometry
  with a dual-core photonic-crystal fiber},}\ }\href {\doibase
  10.1063/1.5024583} {\bibfield  {journal} {\bibinfo  {journal} {Appl. Phys.
  Lett.}\ }\textbf {\bibinfo {volume} {113}},\ \bibinfo {pages} {011112}
  (\bibinfo {year} {2018})}\BibitemShut {NoStop}%
\bibitem [{\citenamefont {Dong}\ \emph {et~al.}(2018)\citenamefont {Dong},
  \citenamefont {Hu}, \citenamefont {Liu}, \citenamefont {Yang}, \citenamefont
  {Wang},\ and\ \citenamefont {Du}}]{dong2018fiber}%
  \BibitemOpen
  \bibfield  {author} {\bibinfo {author} {\bibfnamefont {M.~M.}\ \bibnamefont
  {Dong}}, \bibinfo {author} {\bibfnamefont {Z.~Z.}\ \bibnamefont {Hu}},
  \bibinfo {author} {\bibfnamefont {Y.}~\bibnamefont {Liu}}, \bibinfo {author}
  {\bibfnamefont {B.}~\bibnamefont {Yang}}, \bibinfo {author} {\bibfnamefont
  {Y.~J.}\ \bibnamefont {Wang}}, \ and\ \bibinfo {author} {\bibfnamefont
  {G.~X.}\ \bibnamefont {Du}},\ }\bibfield  {title} {\enquote {\bibinfo {title}
  {A fiber based diamond {RF} b-field sensor and characterization of a small
  helical antenna},}\ }\href {\doibase 10.1063/1.5047495} {\bibfield  {journal}
  {\bibinfo  {journal} {Appl. Phys. Lett.}\ }\textbf {\bibinfo {volume}
  {113}},\ \bibinfo {pages} {131105} (\bibinfo {year} {2018})}\BibitemShut
  {NoStop}%
\bibitem [{\citenamefont {Zhang}\ \emph {et~al.}(2019)\citenamefont {Zhang},
  \citenamefont {Li}, \citenamefont {Du}, \citenamefont {Dong}, \citenamefont
  {Zheng}, \citenamefont {Lin}, \citenamefont {Zhao}, \citenamefont {Zhu},
  \citenamefont {Wang}, \citenamefont {Chen}, \citenamefont {Guo},\ and\
  \citenamefont {Sun}}]{zhang2019thermal}%
  \BibitemOpen
  \bibfield  {author} {\bibinfo {author} {\bibfnamefont {S.-C.}\ \bibnamefont
  {Zhang}}, \bibinfo {author} {\bibfnamefont {S.}~\bibnamefont {Li}}, \bibinfo
  {author} {\bibfnamefont {B.}~\bibnamefont {Du}}, \bibinfo {author}
  {\bibfnamefont {Y.}~\bibnamefont {Dong}}, \bibinfo {author} {\bibfnamefont
  {Y.}~\bibnamefont {Zheng}}, \bibinfo {author} {\bibfnamefont {H.-B.}\
  \bibnamefont {Lin}}, \bibinfo {author} {\bibfnamefont {B.-W.}\ \bibnamefont
  {Zhao}}, \bibinfo {author} {\bibfnamefont {W.}~\bibnamefont {Zhu}}, \bibinfo
  {author} {\bibfnamefont {G.-Z.}\ \bibnamefont {Wang}}, \bibinfo {author}
  {\bibfnamefont {X.-D.}\ \bibnamefont {Chen}}, \bibinfo {author}
  {\bibfnamefont {G.-C.}\ \bibnamefont {Guo}}, \ and\ \bibinfo {author}
  {\bibfnamefont {F.-W.}\ \bibnamefont {Sun}},\ }\bibfield  {title} {\enquote
  {\bibinfo {title} {Thermal-demagnetization-enhanced hybrid fiber-based
  thermometer coupled with nitrogen-vacancy centers},}\ }\href {\doibase
  10.1364/OME.9.004634} {\bibfield  {journal} {\bibinfo  {journal} {Opt. Mater.
  Express}\ }\textbf {\bibinfo {volume} {9}},\ \bibinfo {pages} {4634}
  (\bibinfo {year} {2019})}\BibitemShut {NoStop}%
\bibitem [{\citenamefont {Fang}\ \emph {et~al.}(2013)\citenamefont {Fang},
  \citenamefont {Acosta}, \citenamefont {Santori}, \citenamefont {Huang},
  \citenamefont {Itoh}, \citenamefont {Watanabe}, \citenamefont {Shikata},\
  and\ \citenamefont {Beausoleil}}]{fang2013high}%
  \BibitemOpen
  \bibfield  {author} {\bibinfo {author} {\bibfnamefont {K.}~\bibnamefont
  {Fang}}, \bibinfo {author} {\bibfnamefont {V.~M.}\ \bibnamefont {Acosta}},
  \bibinfo {author} {\bibfnamefont {C.}~\bibnamefont {Santori}}, \bibinfo
  {author} {\bibfnamefont {Z.}~\bibnamefont {Huang}}, \bibinfo {author}
  {\bibfnamefont {K.~M.}\ \bibnamefont {Itoh}}, \bibinfo {author}
  {\bibfnamefont {H.}~\bibnamefont {Watanabe}}, \bibinfo {author}
  {\bibfnamefont {S.}~\bibnamefont {Shikata}}, \ and\ \bibinfo {author}
  {\bibfnamefont {R.~G.}\ \bibnamefont {Beausoleil}},\ }\bibfield  {title}
  {\enquote {\bibinfo {title} {High-sensitivity magnetometry based on quantum
  beats in diamond nitrogen-vacancy centers},}\ }\href {\doibase
  10.1103/PhysRevLett.110.130802} {\bibfield  {journal} {\bibinfo  {journal}
  {Phys. Rev. Lett.}\ }\textbf {\bibinfo {volume} {110}},\ \bibinfo {pages}
  {130802} (\bibinfo {year} {2013})}\BibitemShut {NoStop}%
\bibitem [{\citenamefont {Wang}\ \emph {et~al.}(2015)\citenamefont {Wang},
  \citenamefont {Feng}, \citenamefont {Zhang}, \citenamefont {Chen},
  \citenamefont {Zheng}, \citenamefont {Guo}, \citenamefont {Zhang},
  \citenamefont {Song}, \citenamefont {Guo}, \citenamefont {Fan}, \citenamefont
  {Zou}, \citenamefont {Lou}, \citenamefont {Zhu},\ and\ \citenamefont
  {Wang}}]{wang2015highsensitivity}%
  \BibitemOpen
  \bibfield  {author} {\bibinfo {author} {\bibfnamefont {J.}~\bibnamefont
  {Wang}}, \bibinfo {author} {\bibfnamefont {F.}~\bibnamefont {Feng}}, \bibinfo
  {author} {\bibfnamefont {J.}~\bibnamefont {Zhang}}, \bibinfo {author}
  {\bibfnamefont {J.}~\bibnamefont {Chen}}, \bibinfo {author} {\bibfnamefont
  {Z.}~\bibnamefont {Zheng}}, \bibinfo {author} {\bibfnamefont
  {L.}~\bibnamefont {Guo}}, \bibinfo {author} {\bibfnamefont {W.}~\bibnamefont
  {Zhang}}, \bibinfo {author} {\bibfnamefont {X.}~\bibnamefont {Song}},
  \bibinfo {author} {\bibfnamefont {G.}~\bibnamefont {Guo}}, \bibinfo {author}
  {\bibfnamefont {L.}~\bibnamefont {Fan}}, \bibinfo {author} {\bibfnamefont
  {C.}~\bibnamefont {Zou}}, \bibinfo {author} {\bibfnamefont {L.}~\bibnamefont
  {Lou}}, \bibinfo {author} {\bibfnamefont {W.}~\bibnamefont {Zhu}}, \ and\
  \bibinfo {author} {\bibfnamefont {G.}~\bibnamefont {Wang}},\ }\bibfield
  {title} {\enquote {\bibinfo {title} {High-sensitivity temperature sensing
  using an implanted single nitrogen-vacancy center array in diamond},}\ }\href
  {\doibase 10.1103/PhysRevB.91.155404} {\bibfield  {journal} {\bibinfo
  {journal} {Phys. Rev. B}\ }\textbf {\bibinfo {volume} {91}},\ \bibinfo
  {pages} {155404} (\bibinfo {year} {2015})}\BibitemShut {NoStop}%
\bibitem [{\citenamefont {Wojciechowski}\ \emph {et~al.}(2018)\citenamefont
  {Wojciechowski}, \citenamefont {Karadas}, \citenamefont {Osterkamp},
  \citenamefont {Jankuhn}, \citenamefont {Meijer}, \citenamefont {Jelezko},
  \citenamefont {Huck},\ and\ \citenamefont
  {Andersen}}]{wojciechowski2018precision}%
  \BibitemOpen
  \bibfield  {author} {\bibinfo {author} {\bibfnamefont {A.~M.}\ \bibnamefont
  {Wojciechowski}}, \bibinfo {author} {\bibfnamefont {M.}~\bibnamefont
  {Karadas}}, \bibinfo {author} {\bibfnamefont {C.}~\bibnamefont {Osterkamp}},
  \bibinfo {author} {\bibfnamefont {S.}~\bibnamefont {Jankuhn}}, \bibinfo
  {author} {\bibfnamefont {J.}~\bibnamefont {Meijer}}, \bibinfo {author}
  {\bibfnamefont {F.}~\bibnamefont {Jelezko}}, \bibinfo {author} {\bibfnamefont
  {A.}~\bibnamefont {Huck}}, \ and\ \bibinfo {author} {\bibfnamefont {U.~L.}\
  \bibnamefont {Andersen}},\ }\bibfield  {title} {\enquote {\bibinfo {title}
  {Precision temperature sensing in the presence of magnetic field noise and
  vice-versa using nitrogen-vacancy centers in diamond},}\ }\href {\doibase
  10.1063/1.5026678} {\bibfield  {journal} {\bibinfo  {journal} {Appl. Phys.
  Lett.}\ }\textbf {\bibinfo {volume} {113}},\ \bibinfo {pages} {013502}
  (\bibinfo {year} {2018})}\BibitemShut {NoStop}%
\bibitem [{\citenamefont {Matsuzaki}\ \emph {et~al.}(2016)\citenamefont
  {Matsuzaki}, \citenamefont {Morishita}, \citenamefont {Shimooka},
  \citenamefont {Tashima}, \citenamefont {Kakuyanagi}, \citenamefont {Semba},
  \citenamefont {Munro}, \citenamefont {Yamaguchi}, \citenamefont {Mizuochi},\
  and\ \citenamefont {Saito}}]{matsuzaki2016optically}%
  \BibitemOpen
  \bibfield  {author} {\bibinfo {author} {\bibfnamefont {Y.}~\bibnamefont
  {Matsuzaki}}, \bibinfo {author} {\bibfnamefont {H.}~\bibnamefont
  {Morishita}}, \bibinfo {author} {\bibfnamefont {T.}~\bibnamefont {Shimooka}},
  \bibinfo {author} {\bibfnamefont {T.}~\bibnamefont {Tashima}}, \bibinfo
  {author} {\bibfnamefont {K.}~\bibnamefont {Kakuyanagi}}, \bibinfo {author}
  {\bibfnamefont {K.}~\bibnamefont {Semba}}, \bibinfo {author} {\bibfnamefont
  {W.~J.}\ \bibnamefont {Munro}}, \bibinfo {author} {\bibfnamefont
  {H.}~\bibnamefont {Yamaguchi}}, \bibinfo {author} {\bibfnamefont
  {N.}~\bibnamefont {Mizuochi}}, \ and\ \bibinfo {author} {\bibfnamefont
  {S.}~\bibnamefont {Saito}},\ }\bibfield  {title} {\enquote {\bibinfo {title}
  {Optically detected magnetic resonance of high-density ensemble of
  {NV}-centers in diamond},}\ }\href {\doibase 10.1088/0953-8984/28/27/275302}
  {\bibfield  {journal} {\bibinfo  {journal} {J. Phys.: Condens. Matter}\
  }\textbf {\bibinfo {volume} {28}},\ \bibinfo {pages} {275302} (\bibinfo
  {year} {2016})}\BibitemShut {NoStop}%
\bibitem [{\citenamefont {Mittiga}\ \emph {et~al.}(2018)\citenamefont
  {Mittiga}, \citenamefont {Hsieh}, \citenamefont {Zu}, \citenamefont {Kobrin},
  \citenamefont {Machado}, \citenamefont {Bhattacharyya}, \citenamefont {Rui},
  \citenamefont {Jarmola}, \citenamefont {Choi}, \citenamefont {Budker},\ and\
  \citenamefont {Yao}}]{mittiga2018imaging}%
  \BibitemOpen
  \bibfield  {author} {\bibinfo {author} {\bibfnamefont {T.}~\bibnamefont
  {Mittiga}}, \bibinfo {author} {\bibfnamefont {S.}~\bibnamefont {Hsieh}},
  \bibinfo {author} {\bibfnamefont {C.}~\bibnamefont {Zu}}, \bibinfo {author}
  {\bibfnamefont {B.}~\bibnamefont {Kobrin}}, \bibinfo {author} {\bibfnamefont
  {F.}~\bibnamefont {Machado}}, \bibinfo {author} {\bibfnamefont
  {P.}~\bibnamefont {Bhattacharyya}}, \bibinfo {author} {\bibfnamefont
  {N.}~\bibnamefont {Rui}}, \bibinfo {author} {\bibfnamefont {A.}~\bibnamefont
  {Jarmola}}, \bibinfo {author} {\bibfnamefont {S.}~\bibnamefont {Choi}},
  \bibinfo {author} {\bibfnamefont {D.}~\bibnamefont {Budker}}, \ and\ \bibinfo
  {author} {\bibfnamefont {N.}~\bibnamefont {Yao}},\ }\bibfield  {title}
  {\enquote {\bibinfo {title} {Imaging the local charge environment of
  nitrogen-vacancy centers in diamond},}\ }\href
  {https://link.aps.org/doi/10.1103/PhysRevLett.121.246402} {\bibfield
  {journal} {\bibinfo  {journal} {Phys. Rev. Lett.}\ }\textbf {\bibinfo
  {volume} {121}},\ \bibinfo {pages} {246402} (\bibinfo {year}
  {2018})}\BibitemShut {NoStop}%
\bibitem [{\citenamefont {Hayashi}\ \emph {et~al.}(2018)\citenamefont
  {Hayashi}, \citenamefont {Matsuzaki}, \citenamefont {Taniguchi},
  \citenamefont {Shimo-Oka}, \citenamefont {Nakamura}, \citenamefont {Onoda},
  \citenamefont {Ohshima}, \citenamefont {Morishita}, \citenamefont {Fujiwara},
  \citenamefont {Saito},\ and\ \citenamefont
  {Mizuochi}}]{hayashi2018optimization}%
  \BibitemOpen
  \bibfield  {author} {\bibinfo {author} {\bibfnamefont {K.}~\bibnamefont
  {Hayashi}}, \bibinfo {author} {\bibfnamefont {Y.}~\bibnamefont {Matsuzaki}},
  \bibinfo {author} {\bibfnamefont {T.}~\bibnamefont {Taniguchi}}, \bibinfo
  {author} {\bibfnamefont {T.}~\bibnamefont {Shimo-Oka}}, \bibinfo {author}
  {\bibfnamefont {I.}~\bibnamefont {Nakamura}}, \bibinfo {author}
  {\bibfnamefont {S.}~\bibnamefont {Onoda}}, \bibinfo {author} {\bibfnamefont
  {T.}~\bibnamefont {Ohshima}}, \bibinfo {author} {\bibfnamefont
  {H.}~\bibnamefont {Morishita}}, \bibinfo {author} {\bibfnamefont
  {M.}~\bibnamefont {Fujiwara}}, \bibinfo {author} {\bibfnamefont
  {S.}~\bibnamefont {Saito}}, \ and\ \bibinfo {author} {\bibfnamefont
  {N.}~\bibnamefont {Mizuochi}},\ }\bibfield  {title} {\enquote {\bibinfo
  {title} {Optimization of temperature sensitivity using the optically detected
  magnetic-resonance spectrum of a nitrogen-vacancy center ensemble},}\ }\href
  {\doibase 10.1103/PhysRevApplied.10.034009} {\bibfield  {journal} {\bibinfo
  {journal} {Phys. Rev. Appl.}\ }\textbf {\bibinfo {volume} {10}},\ \bibinfo
  {pages} {034009} (\bibinfo {year} {2018})}\BibitemShut {NoStop}%
\bibitem [{\citenamefont {Ahmadi}\ \emph {et~al.}(2017)\citenamefont {Ahmadi},
  \citenamefont {El-Ella}, \citenamefont {Hansen}, \citenamefont {Huck},\ and\
  \citenamefont {Andersen}}]{ahmadi2017enhanced}%
  \BibitemOpen
  \bibfield  {author} {\bibinfo {author} {\bibfnamefont {S.}~\bibnamefont
  {Ahmadi}}, \bibinfo {author} {\bibfnamefont {H.~A.}\ \bibnamefont {El-Ella}},
  \bibinfo {author} {\bibfnamefont {J.~O.}\ \bibnamefont {Hansen}}, \bibinfo
  {author} {\bibfnamefont {A.}~\bibnamefont {Huck}}, \ and\ \bibinfo {author}
  {\bibfnamefont {U.~L.}\ \bibnamefont {Andersen}},\ }\bibfield  {title}
  {\enquote {\bibinfo {title} {Pump-enhanced continuous-wave magnetometry using
  nitrogen-vacancy ensembles},}\ }\href {\doibase
  10.1103/PhysRevApplied.8.034001} {\bibfield  {journal} {\bibinfo  {journal}
  {Phys. Rev. Appl.}\ }\textbf {\bibinfo {volume} {8}},\ \bibinfo {pages}
  {034001} (\bibinfo {year} {2017})}\BibitemShut {NoStop}%
\bibitem [{\citenamefont {Jensen}\ \emph {et~al.}(2013)\citenamefont {Jensen},
  \citenamefont {Acosta}, \citenamefont {Jarmola},\ and\ \citenamefont
  {Budker}}]{jensen2013light}%
  \BibitemOpen
  \bibfield  {author} {\bibinfo {author} {\bibfnamefont {K.}~\bibnamefont
  {Jensen}}, \bibinfo {author} {\bibfnamefont {V.~M.}\ \bibnamefont {Acosta}},
  \bibinfo {author} {\bibfnamefont {A.}~\bibnamefont {Jarmola}}, \ and\
  \bibinfo {author} {\bibfnamefont {D.}~\bibnamefont {Budker}},\ }\bibfield
  {title} {\enquote {\bibinfo {title} {Light narrowing of magnetic resonances
  in ensembles of nitrogen-vacancy centers in diamond},}\ }\href {\doibase
  10.1103/PhysRevB.87.014115} {\bibfield  {journal} {\bibinfo  {journal} {Phys.
  Rev. B}\ }\textbf {\bibinfo {volume} {87}},\ \bibinfo {pages} {014115}
  (\bibinfo {year} {2013})}\BibitemShut {NoStop}%
\bibitem [{\citenamefont {Acosta}\ \emph {et~al.}(2010)\citenamefont {Acosta},
  \citenamefont {Bauch}, \citenamefont {Ledbetter}, \citenamefont {Waxman},
  \citenamefont {Bouchard},\ and\ \citenamefont
  {Budker}}]{acosta2010temperature}%
  \BibitemOpen
  \bibfield  {author} {\bibinfo {author} {\bibfnamefont {V.~M.}\ \bibnamefont
  {Acosta}}, \bibinfo {author} {\bibfnamefont {E.}~\bibnamefont {Bauch}},
  \bibinfo {author} {\bibfnamefont {M.~P.}\ \bibnamefont {Ledbetter}}, \bibinfo
  {author} {\bibfnamefont {A.}~\bibnamefont {Waxman}}, \bibinfo {author}
  {\bibfnamefont {L.-S.}\ \bibnamefont {Bouchard}}, \ and\ \bibinfo {author}
  {\bibfnamefont {D.}~\bibnamefont {Budker}},\ }\bibfield  {title} {\enquote
  {\bibinfo {title} {Temperature dependence of the nitrogen-vacancy magnetic
  resonance in diamond},}\ }\href
  {https://journals.aps.org/prl/abstract/10.1103/PhysRevLett.104.070801}
  {\bibfield  {journal} {\bibinfo  {journal} {Phys. Rev. Lett.}\ }\textbf
  {\bibinfo {volume} {104}},\ \bibinfo {pages} {070801} (\bibinfo {year}
  {2010})}\BibitemShut {NoStop}%
\bibitem [{\citenamefont {Li}\ \emph {et~al.}(2017)\citenamefont {Li},
  \citenamefont {Gong}, \citenamefont {Chen}, \citenamefont {Li}, \citenamefont
  {Zhao}, \citenamefont {Dong}, \citenamefont {Guo},\ and\ \citenamefont
  {Sun}}]{li2017temperature}%
  \BibitemOpen
  \bibfield  {author} {\bibinfo {author} {\bibfnamefont {C.-C.}\ \bibnamefont
  {Li}}, \bibinfo {author} {\bibfnamefont {M.}~\bibnamefont {Gong}}, \bibinfo
  {author} {\bibfnamefont {X.-D.}\ \bibnamefont {Chen}}, \bibinfo {author}
  {\bibfnamefont {S.}~\bibnamefont {Li}}, \bibinfo {author} {\bibfnamefont
  {B.-W.}\ \bibnamefont {Zhao}}, \bibinfo {author} {\bibfnamefont
  {Y.}~\bibnamefont {Dong}}, \bibinfo {author} {\bibfnamefont {G.-C.}\
  \bibnamefont {Guo}}, \ and\ \bibinfo {author} {\bibfnamefont {F.-W.}\
  \bibnamefont {Sun}},\ }\bibfield  {title} {\enquote {\bibinfo {title}
  {Temperature dependent energy gap shifts of single color center in diamond
  based on modified varshni equation},}\ }\href {\doibase
  10.1016/j.diamond.2017.03.002} {\bibfield  {journal} {\bibinfo  {journal}
  {Diam. Relat. Mater.}\ }\textbf {\bibinfo {volume} {74}},\ \bibinfo {pages}
  {119} (\bibinfo {year} {2017})}\BibitemShut {NoStop}%
\bibitem [{\citenamefont {Doherty}\ \emph {et~al.}(2013)\citenamefont
  {Doherty}, \citenamefont {Manson}, \citenamefont {Delaney}, \citenamefont
  {Jelezko}, \citenamefont {Wrachtrup},\ and\ \citenamefont
  {Hollenberg}}]{noauthor2013nitrogen}%
  \BibitemOpen
  \bibfield  {author} {\bibinfo {author} {\bibfnamefont {M.~W.}\ \bibnamefont
  {Doherty}}, \bibinfo {author} {\bibfnamefont {N.~B.}\ \bibnamefont {Manson}},
  \bibinfo {author} {\bibfnamefont {P.}~\bibnamefont {Delaney}}, \bibinfo
  {author} {\bibfnamefont {F.}~\bibnamefont {Jelezko}}, \bibinfo {author}
  {\bibfnamefont {J.}~\bibnamefont {Wrachtrup}}, \ and\ \bibinfo {author}
  {\bibfnamefont {L.~C.}\ \bibnamefont {Hollenberg}},\ }\bibfield  {title}
  {\enquote {\bibinfo {title} {The nitrogen-vacancy colour centre in
  diamond},}\ }\href {\doibase 10.1016/j.physrep.2013.02.001} {\bibfield
  {journal} {\bibinfo  {journal} {Phys. Rep.}\ }\textbf {\bibinfo {volume}
  {528}},\ \bibinfo {pages} {1} (\bibinfo {year} {2013})}\BibitemShut {NoStop}%
\bibitem [{\citenamefont {Doherty}\ \emph {et~al.}(2012)\citenamefont
  {Doherty}, \citenamefont {Dolde}, \citenamefont {Fedder}, \citenamefont
  {Jelezko}, \citenamefont {Wrachtrup}, \citenamefont {Manson},\ and\
  \citenamefont {Hollenberg}}]{doherty2012theory}%
  \BibitemOpen
  \bibfield  {author} {\bibinfo {author} {\bibfnamefont {M.~W.}\ \bibnamefont
  {Doherty}}, \bibinfo {author} {\bibfnamefont {F.}~\bibnamefont {Dolde}},
  \bibinfo {author} {\bibfnamefont {H.}~\bibnamefont {Fedder}}, \bibinfo
  {author} {\bibfnamefont {F.}~\bibnamefont {Jelezko}}, \bibinfo {author}
  {\bibfnamefont {J.}~\bibnamefont {Wrachtrup}}, \bibinfo {author}
  {\bibfnamefont {N.~B.}\ \bibnamefont {Manson}}, \ and\ \bibinfo {author}
  {\bibfnamefont {L.~C.~L.}\ \bibnamefont {Hollenberg}},\ }\bibfield  {title}
  {\enquote {\bibinfo {title} {Theory of the ground-state spin of the {NV}$^-$
  in diamond},}\ }\href {\doibase 10.1103/PhysRevB.85.205203} {\bibfield
  {journal} {\bibinfo  {journal} {Phys. Rev. B}\ }\textbf {\bibinfo {volume}
  {85}},\ \bibinfo {pages} {205203} (\bibinfo {year} {2012})}\BibitemShut
  {NoStop}%
\bibitem [{\citenamefont {El-Ella}\ \emph {et~al.}(2017)\citenamefont
  {El-Ella}, \citenamefont {Ahmadi}, \citenamefont {Wojciechowski},
  \citenamefont {Huck},\ and\ \citenamefont {Andersen}}]{ella2017optimised}%
  \BibitemOpen
  \bibfield  {author} {\bibinfo {author} {\bibfnamefont {H.~A.~R.}\
  \bibnamefont {El-Ella}}, \bibinfo {author} {\bibfnamefont {S.}~\bibnamefont
  {Ahmadi}}, \bibinfo {author} {\bibfnamefont {A.~M.}\ \bibnamefont
  {Wojciechowski}}, \bibinfo {author} {\bibfnamefont {A.}~\bibnamefont {Huck}},
  \ and\ \bibinfo {author} {\bibfnamefont {U.~L.}\ \bibnamefont {Andersen}},\
  }\bibfield  {title} {\enquote {\bibinfo {title} {Optimised frequency
  modulation for continuous-wave optical magnetic resonance sensing using
  nitrogen-vacancy ensembles},}\ }\href {\doibase 10.1364/OE.25.014809}
  {\bibfield  {journal} {\bibinfo  {journal} {Opt. Express}\ }\textbf {\bibinfo
  {volume} {25}},\ \bibinfo {pages} {14809} (\bibinfo {year}
  {2017})}\BibitemShut {NoStop}%
\bibitem [{\citenamefont {Clevenson}\ \emph {et~al.}(2018)\citenamefont
  {Clevenson}, \citenamefont {Pham}, \citenamefont {Teale}, \citenamefont
  {Johnson}, \citenamefont {Englund},\ and\ \citenamefont
  {Braje}}]{clevenson2018robust}%
  \BibitemOpen
  \bibfield  {author} {\bibinfo {author} {\bibfnamefont {H.}~\bibnamefont
  {Clevenson}}, \bibinfo {author} {\bibfnamefont {L.~M.}\ \bibnamefont {Pham}},
  \bibinfo {author} {\bibfnamefont {C.}~\bibnamefont {Teale}}, \bibinfo
  {author} {\bibfnamefont {K.}~\bibnamefont {Johnson}}, \bibinfo {author}
  {\bibfnamefont {D.}~\bibnamefont {Englund}}, \ and\ \bibinfo {author}
  {\bibfnamefont {D.}~\bibnamefont {Braje}},\ }\bibfield  {title} {\enquote
  {\bibinfo {title} {Robust high-dynamic-range vector magnetometry with
  nitrogen-vacancy centers in diamond},}\ }\href {\doibase 10.1063/1.5034216}
  {\bibfield  {journal} {\bibinfo  {journal} {Appl. Phys. Lett.}\ }\textbf
  {\bibinfo {volume} {112}},\ \bibinfo {pages} {252406} (\bibinfo {year}
  {2018})}\BibitemShut {NoStop}%
\bibitem [{\citenamefont {Schoenfeld}\ and\ \citenamefont
  {Harneit}(2011)}]{schoenfeld2011real}%
  \BibitemOpen
  \bibfield  {author} {\bibinfo {author} {\bibfnamefont {R.~S.}\ \bibnamefont
  {Schoenfeld}}\ and\ \bibinfo {author} {\bibfnamefont {W.}~\bibnamefont
  {Harneit}},\ }\bibfield  {title} {\enquote {\bibinfo {title} {Real time
  magnetic field sensing and imaging using a single spin in diamond},}\ }\href
  {\doibase 10.1103/PhysRevLett.106.030802} {\bibfield  {journal} {\bibinfo
  {journal} {Phys. Rev. Lett.}\ }\textbf {\bibinfo {volume} {106}},\ \bibinfo
  {pages} {030802} (\bibinfo {year} {2011})}\BibitemShut {NoStop}%
\bibitem [{\citenamefont {Shao}\ \emph {et~al.}(2016)\citenamefont {Shao},
  \citenamefont {Zhang}, \citenamefont {Markham}, \citenamefont {Edmonds},\
  and\ \citenamefont {Loncar}}]{shao2016diamond}%
  \BibitemOpen
  \bibfield  {author} {\bibinfo {author} {\bibfnamefont {L.}~\bibnamefont
  {Shao}}, \bibinfo {author} {\bibfnamefont {M.}~\bibnamefont {Zhang}},
  \bibinfo {author} {\bibfnamefont {M.}~\bibnamefont {Markham}}, \bibinfo
  {author} {\bibfnamefont {A.~M.}\ \bibnamefont {Edmonds}}, \ and\ \bibinfo
  {author} {\bibfnamefont {M.}~\bibnamefont {Loncar}},\ }\bibfield  {title}
  {\enquote {\bibinfo {title} {Diamond radio receiver: Nitrogen-vacancy centers
  as fluorescent transducers of microwave signals},}\ }\href {\doibase
  10.1103/PhysRevApplied.6.064008} {\bibfield  {journal} {\bibinfo  {journal}
  {Phys. Rev. Appl.}\ }\textbf {\bibinfo {volume} {6}},\ \bibinfo {pages}
  {064008} (\bibinfo {year} {2016})}\BibitemShut {NoStop}%
\bibitem [{\citenamefont {Schloss}\ \emph {et~al.}(2018)\citenamefont
  {Schloss}, \citenamefont {Barry}, \citenamefont {Turner},\ and\ \citenamefont
  {Walsworth}}]{schloss2018simultaneous}%
  \BibitemOpen
  \bibfield  {author} {\bibinfo {author} {\bibfnamefont {J.~M.}\ \bibnamefont
  {Schloss}}, \bibinfo {author} {\bibfnamefont {J.~F.}\ \bibnamefont {Barry}},
  \bibinfo {author} {\bibfnamefont {M.~J.}\ \bibnamefont {Turner}}, \ and\
  \bibinfo {author} {\bibfnamefont {R.~L.}\ \bibnamefont {Walsworth}},\
  }\bibfield  {title} {\enquote {\bibinfo {title} {Simultaneous broadband
  vector magnetometry using solid-state spins},}\ }\href {\doibase
  10.1103/PhysRevApplied.10.034044} {\bibfield  {journal} {\bibinfo  {journal}
  {Phys. Rev. Appl.}\ }\textbf {\bibinfo {volume} {10}},\ \bibinfo {pages}
  {034044} (\bibinfo {year} {2018})}\BibitemShut {NoStop}%
\bibitem [{\citenamefont {Kitazawa}\ \emph {et~al.}(2017)\citenamefont
  {Kitazawa}, \citenamefont {Matsuzaki}, \citenamefont {Saijo}, \citenamefont
  {Kakuyanagi}, \citenamefont {Saito},\ and\ \citenamefont
  {Ishi-Hayase}}]{kitazawa2017vector}%
  \BibitemOpen
  \bibfield  {author} {\bibinfo {author} {\bibfnamefont {S.}~\bibnamefont
  {Kitazawa}}, \bibinfo {author} {\bibfnamefont {Y.}~\bibnamefont {Matsuzaki}},
  \bibinfo {author} {\bibfnamefont {S.}~\bibnamefont {Saijo}}, \bibinfo
  {author} {\bibfnamefont {K.}~\bibnamefont {Kakuyanagi}}, \bibinfo {author}
  {\bibfnamefont {S.}~\bibnamefont {Saito}}, \ and\ \bibinfo {author}
  {\bibfnamefont {J.}~\bibnamefont {Ishi-Hayase}},\ }\bibfield  {title}
  {\enquote {\bibinfo {title} {Vector-magnetic-field sensing via multifrequency
  control of nitrogen-vacancy centers in diamond},}\ }\href {\doibase
  10.1103/PhysRevA.96.042115} {\bibfield  {journal} {\bibinfo  {journal} {Phys.
  Rev. A}\ }\textbf {\bibinfo {volume} {96}},\ \bibinfo {pages} {042115}
  (\bibinfo {year} {2017})}\BibitemShut {NoStop}%
\bibitem [{\citenamefont {Yahata}\ \emph {et~al.}(2019)\citenamefont {Yahata},
  \citenamefont {Matsuzaki}, \citenamefont {Saito}, \citenamefont {Watanabe},\
  and\ \citenamefont {Ishi-Hayase}}]{yahata2019demonstration}%
  \BibitemOpen
  \bibfield  {author} {\bibinfo {author} {\bibfnamefont {K.}~\bibnamefont
  {Yahata}}, \bibinfo {author} {\bibfnamefont {Y.}~\bibnamefont {Matsuzaki}},
  \bibinfo {author} {\bibfnamefont {S.}~\bibnamefont {Saito}}, \bibinfo
  {author} {\bibfnamefont {H.}~\bibnamefont {Watanabe}}, \ and\ \bibinfo
  {author} {\bibfnamefont {J.}~\bibnamefont {Ishi-Hayase}},\ }\bibfield
  {title} {\enquote {\bibinfo {title} {Demonstration of vector magnetic field
  sensing by simultaneous control of nitrogen-vacancy centers in diamond using
  multi-frequency microwave pulses},}\ }\href {\doibase 10.1063/1.5079925}
  {\bibfield  {journal} {\bibinfo  {journal} {Appl. Phys. Lett.}\ }\textbf
  {\bibinfo {volume} {114}},\ \bibinfo {pages} {022404} (\bibinfo {year}
  {2019})}\BibitemShut {NoStop}%
\bibitem [{\citenamefont {Farfurnik}\ \emph {et~al.}(2017)\citenamefont
  {Farfurnik}, \citenamefont {Alfasi}, \citenamefont {Masis}, \citenamefont
  {Kauffmann}, \citenamefont {Farchi}, \citenamefont {Romach}, \citenamefont
  {Hovav}, \citenamefont {Buks},\ and\ \citenamefont
  {Bar-Gill}}]{farfurnik2017enhanced}%
  \BibitemOpen
  \bibfield  {author} {\bibinfo {author} {\bibfnamefont {D.}~\bibnamefont
  {Farfurnik}}, \bibinfo {author} {\bibfnamefont {N.}~\bibnamefont {Alfasi}},
  \bibinfo {author} {\bibfnamefont {S.}~\bibnamefont {Masis}}, \bibinfo
  {author} {\bibfnamefont {Y.}~\bibnamefont {Kauffmann}}, \bibinfo {author}
  {\bibfnamefont {E.}~\bibnamefont {Farchi}}, \bibinfo {author} {\bibfnamefont
  {Y.}~\bibnamefont {Romach}}, \bibinfo {author} {\bibfnamefont
  {Y.}~\bibnamefont {Hovav}}, \bibinfo {author} {\bibfnamefont
  {E.}~\bibnamefont {Buks}}, \ and\ \bibinfo {author} {\bibfnamefont
  {N.}~\bibnamefont {Bar-Gill}},\ }\bibfield  {title} {\enquote {\bibinfo
  {title} {Enhanced concentrations of nitrogen-vacancy centers in diamond
  through {TEM} irradiation},}\ }\href {\doibase 10.1063/1.4993257} {\bibfield
  {journal} {\bibinfo  {journal} {Appl. Phys. Lett.}\ }\textbf {\bibinfo
  {volume} {111}},\ \bibinfo {pages} {123101} (\bibinfo {year}
  {2017})}\BibitemShut {NoStop}%
\bibitem [{\citenamefont {{McLellan}}\ \emph {et~al.}(2016)\citenamefont
  {{McLellan}}, \citenamefont {Myers}, \citenamefont {Kraemer}, \citenamefont
  {Ohno}, \citenamefont {Awschalom},\ and\ \citenamefont
  {Bleszynski~Jayich}}]{mclellan2016patterned}%
  \BibitemOpen
  \bibfield  {author} {\bibinfo {author} {\bibfnamefont {C.~A.}\ \bibnamefont
  {{McLellan}}}, \bibinfo {author} {\bibfnamefont {B.~A.}\ \bibnamefont
  {Myers}}, \bibinfo {author} {\bibfnamefont {S.}~\bibnamefont {Kraemer}},
  \bibinfo {author} {\bibfnamefont {K.}~\bibnamefont {Ohno}}, \bibinfo {author}
  {\bibfnamefont {D.~D.}\ \bibnamefont {Awschalom}}, \ and\ \bibinfo {author}
  {\bibfnamefont {A.~C.}\ \bibnamefont {Bleszynski~Jayich}},\ }\bibfield
  {title} {\enquote {\bibinfo {title} {Patterned formation of highly coherent
  nitrogen-vacancy centers using a focused electron irradiation technique},}\
  }\href {\doibase 10.1021/acs.nanolett.5b05304} {\bibfield  {journal}
  {\bibinfo  {journal} {Nano Lett.}\ }\textbf {\bibinfo {volume} {16}},\
  \bibinfo {pages} {2450} (\bibinfo {year} {2016})}\BibitemShut {NoStop}%
\bibitem [{\citenamefont {Eichhorn}\ \emph {et~al.}(2019)\citenamefont
  {Eichhorn}, \citenamefont {{McLellan}},\ and\ \citenamefont
  {Bleszynski~Jayich}}]{eichhorn2019optimizing}%
  \BibitemOpen
  \bibfield  {author} {\bibinfo {author} {\bibfnamefont {T.~R.}\ \bibnamefont
  {Eichhorn}}, \bibinfo {author} {\bibfnamefont {C.~A.}\ \bibnamefont
  {{McLellan}}}, \ and\ \bibinfo {author} {\bibfnamefont {A.~C.}\ \bibnamefont
  {Bleszynski~Jayich}},\ }\bibfield  {title} {\enquote {\bibinfo {title}
  {Optimizing the formation of depth-confined nitrogen vacancy center spin
  ensembles in diamond for quantum sensing},}\ }\href {\doibase
  10.1103/PhysRevMaterials.3.113802} {\bibfield  {journal} {\bibinfo  {journal}
  {Phys. Rev. Mater.}\ }\textbf {\bibinfo {volume} {3}},\ \bibinfo {pages}
  {113802} (\bibinfo {year} {2019})}\BibitemShut {NoStop}%
\bibitem [{\citenamefont {Duan}\ \emph {et~al.}(2018)\citenamefont {Duan},
  \citenamefont {Kavatamane}, \citenamefont {Arumugam}, \citenamefont {Rahane},
  \citenamefont {Tzeng}, \citenamefont {Chang}, \citenamefont {Sumiya},
  \citenamefont {Onoda}, \citenamefont {Isoya},\ and\ \citenamefont
  {Balasubramanian}}]{duan2018enhancing}%
  \BibitemOpen
  \bibfield  {author} {\bibinfo {author} {\bibfnamefont {D.}~\bibnamefont
  {Duan}}, \bibinfo {author} {\bibfnamefont {V.~K.}\ \bibnamefont
  {Kavatamane}}, \bibinfo {author} {\bibfnamefont {S.~R.}\ \bibnamefont
  {Arumugam}}, \bibinfo {author} {\bibfnamefont {G.}~\bibnamefont {Rahane}},
  \bibinfo {author} {\bibfnamefont {Y.-K.}\ \bibnamefont {Tzeng}}, \bibinfo
  {author} {\bibfnamefont {H.-C.}\ \bibnamefont {Chang}}, \bibinfo {author}
  {\bibfnamefont {H.}~\bibnamefont {Sumiya}}, \bibinfo {author} {\bibfnamefont
  {S.}~\bibnamefont {Onoda}}, \bibinfo {author} {\bibfnamefont
  {J.}~\bibnamefont {Isoya}}, \ and\ \bibinfo {author} {\bibfnamefont
  {G.}~\bibnamefont {Balasubramanian}},\ }\bibfield  {title} {\enquote
  {\bibinfo {title} {Enhancing fluorescence excitation and collection from the
  nitrogen-vacancy center in diamond through a micro-concave mirror},}\ }\href
  {\doibase 10.1063/1.5037807} {\bibfield  {journal} {\bibinfo  {journal}
  {Appl. Phys. Lett.}\ }\textbf {\bibinfo {volume} {113}},\ \bibinfo {pages}
  {041107} (\bibinfo {year} {2018})}\BibitemShut {NoStop}%
\end{thebibliography}
%
\end{document}